\documentclass[prb,twocolumn,amsmath,amssymb,floatfix]{revtex4}
\usepackage{amsmath}
\usepackage{amssymb}
\usepackage{graphicx}
\usepackage{dcolumn}
\usepackage{bm}
\usepackage{bm}
\usepackage{blindtext}
\usepackage{natbib}
\usepackage[pdfborder={0 0 0},dvipdfmx,colorlinks=true,linkcolor=blue,urlcolor=blue,citecolor=green]{hyperref}
\newcommand{\be}{\begin{eqnarray}}
\newcommand{\ee}{\end{eqnarray}}

\newcommand{\wbe}{\begin{widetext}}
\newcommand{\wee}{\end{widetext}}
\newcommand{\oncite}{\onlinecite}

\begin{document}

\title{Electron-spin to Phonon Coupling in Graphene Decorated with Heavy Adatoms}

\author{Jhih-Shih You$^{1,2,3}$, Daw-Wei Wang$^{1,2}$,  Miguel A. Cazalilla$^{1,2}$}


\affiliation{$^{1}$ Physics Department and Frontier Research Center on Fundamental and Applied Sciences of Matter, National Tsing-Hua University, Hsinchu,
Taiwan
\\
$^{2}$ Physics Division, National Center for Theoretical Sciences,
Hsinchu, Taiwan
\\
$^{3}$ Department of Physics, University of California, San Diego, CA 92093
}
\begin{abstract}
The naturally weak spin-orbit coupling in Graphene can be largely enhanced by adatom deposition~(e.g.  Weeks \emph{et al.} Phys. Rev. X {\bf 1}, 021001 (2011)). However, the dynamics of the adatoms also induces a coupling between phonons and the electron spin. Using group theory and a tight-binding model, we systematically investigate the coupling between the low-energy in-plane phonons and the electron spin in single-layer graphene uniformly decorated with heavy adatoms. Our results provide the foundation for future investigations of spin transport and  superconductivity in this system. In order to quantify the effect of the coupling  to the lattice on the electronic spin dynamics, we  compute the spin-flip rate of electrons and holes. We show that the latter exhibits a strong dependence on the quasi-particle energy and system temperature.
\end{abstract}
\maketitle
\section{Introduction}

 Two-dimensional topological insulators exhibiting the quantum spin Hall effect~(QSHE) hold great potential  for applications in low-power consumption electronics.~\cite{Kane,Hasan,Qi} However, so
far very few such materials are
available.~\cite{mercury_tell,other}
This makes the quest for new materials exhibiting the QSHE
a very important research  enterprise. Early in the history of the field, it was pointed out by Kane and Mele~\cite{Kane}
that graphene,\cite{novoselov,neto} if endowed with a sizable spin-orbit coupling~(SOC), would exhibit a robust QSHE. However, the strength of the SOC in graphene is just a few tens of micro-volts,~\cite{Hernando,Yao,Min,Gmitra} which makes the QSHE
in graphene extremely fragile and, in practice, inaccessible in actual experiments.

 Nevertheless, after its isolation~\cite{novoselov}  in 2004,  graphene, a monolayer of carbon atoms arranged in a honeycomb lattice, has become an extremely
popular 2D material because it can be cheaply produced in large-size samples displaying large carrier mobilities. Furthermore, the carrier charge and density can be easily tuned by gating.~\cite{neto} In recently years, several methods for enhancing
the SOC in graphene have been put forward.~\cite{neto2,Samir,Qiao,Alicea1} They largely rely on
adatom deposition, and can lead to either predominantly
Rashba-type SOC,~\cite{neto2} predominantly intrinsic-type SOC,~\cite{Alicea1} or both,~\cite{Samir,Qiao} depending on the adatom
type.  Castro Neto and Guinea~\cite{neto2} pointed out that chemisorption of atomic species like Hydrogen or Fluorine, which bind to the carbon atoms by inducing $sp^{2}$-$sp^{3}$ hybridization,  locally induces a spin orbit coupling of the Rashba-type, which can be as large as several tens of meV.
This proposal has been recently verified experimentally
by Balakrishnan and coworkers,~\cite{natphys1} who measured
the spin-Hall effect~(SHE) in lightly hydrogenated graphene devices.

 In addition, it was proposed by Weeks et al.~\cite{Alicea1} that, by  deposition  of certain heavy adatoms like Indium and Thallium, the strength of the intrinsic-type SOC can be largely enhanced in single-layer graphene. The enhancement is due to second-neighbor hopping that is mediated by the spin-orbit-splitted $p$-orbitals of the adatoms, which strongly hybridize with  the unoccupied $\pi$-levels of graphene. Such an enhancement of the intrinsic SOC can lead to a band gap of the order of several tens of meV, and therefore, to a much more robust QSHE. Furthermore, it was noted that, for these atomic species, the adatoms find their lowest absorption energies at the center of the hexagons
on the honeycomb lattice~(the so-called H position). Indeed,  first principle calculations by Weeks and coworkers~\cite{Alicea1} found that the heavy adatoms at the  H position have binding energies of  $-0.525$ eV, being the energy difference between H and
B position (i.e. the bridge between two neighbor carbon atoms) $80$ meV. In this configuration,  the Rashba-type SOC, which is  detrimental for QSHE~\cite{Kane} is entirely
absent at the $K$ and $K^{\prime}$ points (but grows linearly with the crystal-momentum separation to those points).
Deposition of other atomic species, such as transition metal elements  like Iridium and Osmium, has been also studied.~\cite{Jun} In this case,  the details of  microscopic mechanism by which SOC is induced on the graphene layer by  transition-metal atoms containing $d$ orbitals is different from the atomic species studied by Weeks and coworkers,~\cite{Alicea1} for which the $p$-orbitals play the dominant role.

 It is worth pointing out that recent experiments~\cite{jaja2} carried both in CVD graphene (where Copper atoms exist in the form of a residue resulting from the fabrication process) and in \emph{intentionally} decorated graphene with  noble metal clusters of e.g. Cu, and Au, support the idea that decoration can enhance the SOC in graphene. As in the case of Hydrogen,~\cite{natphys1} decoration produces a robust SHE in
standard Hall-bar devices made with CVD and exfoliated graphene samples that have been decorated with adatoms. Indeed, single-layer graphene is particularly prone to the existence of resonances in the
neighborhood of the Dirac point induced by adatoms (see~\cite{Katsnelson} and references therein). This has been recently shown  to lead to a sizable enhancement  the SHE.~\cite{prl2014,jaja2}

 Nevertheless, the possibility of a proximity-induced SOC in graphene still remains controversial, as a recent  experiment~\cite{Yu2014} found no experimental evidence of such effect induced  in  the  magnetoresistance, quantum Hall effect and non-local  spin Hall effect of graphene decorated with Indium. However, other studies~\cite{Rashba,Hector}
have found evidence of such proximity induced SOC by Gold~\cite{Rashba}
and Lead~\cite{Hector} in the
quasi-particle spectrum measured either through angle-resolved
phono-emission~\cite{Rashba} or scanning tunneling spectroscopy.~\cite{Hector} Further studies are therefore needed in order to
better clarify the complexities of the coupling between the
adatoms and the electrons on the graphene layer.

One important concern with the decoration of graphene by heavy adatoms is that the latter are often physisorpbed rather than chemisorpbed. Thus, the binding of the adatoms to the layer is rather weak and a native estimation  of the phonon frequencies of the modes associated with the motion of the adatom yields a relatively low characteristic energy of $\approx 10$ meV.~\footnote{This estimation is made using the difference in calculated~\cite{Alicea1}
absorption energies at the two most stable positions, namely the H (i.e. center of the carbon hexagon) and B (i.e. middle side of the hexagon) positions.}  Since the adatom is responsible for the SOC, when its position fluctuates, it will affect the electron spin. Added to the relatively low-energy of the relevant phonons, it can be expected that the lattice dynamics in adatom decorated graphene
can have a substantial effect on the spin-dynamics of the low-energy electrons and holes. The precise way in which those low-energy phonons couple to the electron spin degree of freedom is the subject of this work. In addition, in order to assess the effect of this coupling on the spin-dynamics,  we have computed the phonon-contribution to the spin-flip rate of electrons and holes.

 The rest paper is organized as follows. In Sec.~\ref{sec:symmetry}, we use symmetry arguments as well as a simple model to obtain the phonon spectrum of decorated graphene. In Sec.~\ref{es}, we review the most important facts about the tight-biding model for the electronic structure of single-layer graphene decorated with a periodic array of heavy atoms sitting on the H position introduced in Ref.~[\onlinecite{Alicea1}]. In Sec.~\ref{sec:coupling}, the coupling  of the low-energy phonons to the electron spin is obtained both from symmetry arguments and the tight-biding model. The
consequences of this coupling for the electron spin-flip
rate are discussed in Sec.~\ref{sec:spinres}. Finally,  we discuss our results in Sec.~\ref{sec:Discussion} and summarize our main conclusions  in Sec.~\ref{sec:Summary}. Some of the more technical details of the calculations have been relegated to the Appendices.

\section{Phonon spectrum of adatom decorated graphene}\label{sec:symmetry}

 In order to obtain the spectral degeneracies of the
phonon spectrum around special points of the Brillouin zone~(BZ), we first rely upon group theory. Thus we analyze the symmetry properties of graphene covered by a periodic layer of
adatoms of the type considered by Weeks  \emph{et al.}~\cite{Alicea1} The symmetry group  constrains the types of phonon-mediated spin-orbit coupling. Since the adatoms sit periodically at the H position on top of the graphene layer~(see Fig. 1, for example), each unit cell contains two carbon atoms and one adatom. Note that, in this configuration, the mirror symmetry where $z\rightarrow-z$ is absent. Therefore, the point group of graphene with adatoms on the H position is $C_{6v}$, which has $12$ elements, whose equivalence classes are $\{E,2 C_6,2 C_3,C_2, 3 \sigma_a,3 \sigma'_a\}$ where $C_m$ describes a rotation by $2\pi/m$  by an axis~($z$) perpendicular to the graphene layer, $\sigma_a$ describes reflections containing the $z$ axis that leave the A and B sublattices invariant, and $\sigma'_a$ describes reflections that swap the two sublattices.~\cite{Basko}

\begin{figure}[htbp!]
\centering
\includegraphics[width=8cm]{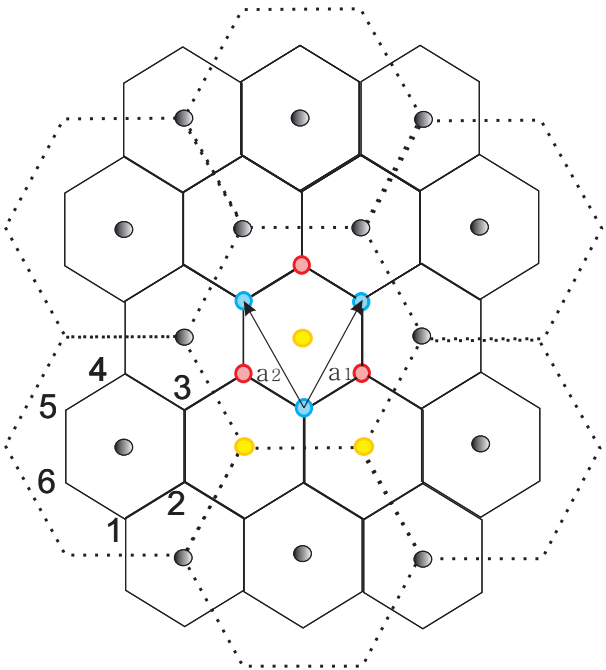}
\caption{Enlarged unit cell contains  $9$ atoms, $6$ of which are carbon (in red and blue, for $A$ and $B$ sublattices, respectively) plus $3$ adatoms (in yellow). $a_1$ and $a_2$ are two vectors of unit cell.
}\label{fig_group}
\end{figure}

 Generally speaking, two kinds of phonon modes affect the dynamics of electrons at low energies. Phonons near  the $\Gamma$  point can scatter low-energy electrons within the same valley. In addition, phonons at $K (K^{\prime})$ points can also scatter electrons between valleys because ${\bf K}^{\prime}$ is equivalent to $-{\bf K}$ and $3\mathbf{K}$ is a reciprocal lattice vector \emph{umklapp} scattering (i.e. $(\mathbf{K})_{\mathrm{e}}+(\mathbf{K})_{\mathrm{ph}}+\mathbf{G}=(\mathbf{K}^{\prime})_{\mathrm{e}^{\prime}}$ for $\mathbf{G}= -3\mathbf{K}$). In order to treat the inter- and intra-valley phonon scattering at the same time, it is convenient to follow  Ref.~[\onlinecite{Basko}] and  triple the unit cell (see Fig.~\ref{fig_group}), which means that two translations $t_{\boldsymbol{a}1}$ and $t_{\boldsymbol{a}2}$ are factored out from translation group and must be included in the point group $C_{6v}$. This yields the group $C^{\prime\prime}_{6v}=C_{6v}+t_{\boldsymbol{a}1}C_{6v}+t_{\boldsymbol{a}2}C_{6v}$.  This approach maps the $K$ and $K^{\prime}$ points to the $\Gamma$ point
of a lattice with a unit cell that is three times larger. Thus, the states at  $K, K^{\prime}$ become degenerate and we can deal with them simultaneously. Therefore, the possible inter- and intra-valley couplings between the phonons and the electron spin, which in addition must respect time-reversal symmetry, can be  classified according to the \emph{irreducible} representations (\emph{irreps})
of the $C''_{6v}$.
\subsection{Symmetry analysis of the phonon modes}

First, let us study the classification of in-plane phonon modes according to the symmetry operations of $C''
_{6v}$. The tripled unit cell contains $9$ atoms, i.e. $6$ carbons and $3$ adatoms. When one atom is mapped onto an equivalent atom by a symmetry operation in  $C''_6$, it contributes 1 to the character of the  transformation. For example, under $C_3$, only the $3$ adatoms are invariant, and therefore the character for this element equals $3$. The character table for the different symmetry transformations in the tripled-unit cell lattice is shown in Table~\ref{c6prime}. By applying the orthogonality theorem, this representation can be reduced to $2A_1+B_1+E'_1+G'$.
\begin{table}[t]
\centering
\begin{tabular}{|c|c|c|c|c|c|c|c|c|c|}\hline
$C''
_{6v}$& $E$ & $2 t_{a1}$ & $3 C_2$ & $2 C_3$ & $4 t_{a1} C_3$ &$6 C_6$ & $3 \sigma'_a$ &$6 t_{a1}\sigma'_a$ & $9\sigma_a$ \\ \hline
$\chi$& 9  & 0   & 1        &3         &3&1&3&0&3 \\\hline
\end{tabular}
\caption{Character table for the vibrational representation of the group $C''_{6v}$}\label{c6prime}
\end{table}
Taking into account that a two dimensional vector like the atom displacement $\boldsymbol{u} = (x, y)$ transforms according to the $E_1$ \emph{irrep} of $C_{6v}$,  the vector of in-plane vibrations belongs to an $18$-dimensional \emph{reducible} representation. The latter can be decomposed into
$(2A_1+B_1+E'_1+G')\times E_1 = 2E_1+E_2+E'_1+E'_2+2G'$,
where we have used that~\cite{Basko} $A_1\times E_1=E_1, B_1\times E_1=E_2,E'_1\times E_1=G', G'\times E_1=E'_1+E'_2+G'$;
$G'$ is a four-dimensional \emph{irrep} and the $E_1, E_2, E^{\prime}_{1},E^{\prime}_2$  are two dimensional \emph{irreps}.

In order to better identify which of the above modes belong to the $\Gamma$ point of the BZ, we can go back to the original three-atom unit cell rather than using the tripled unit cell. The original unit cell transforms according to the $C_{6v}$ group. This group is also the little group for the $\Gamma$ point (the little
group for  $K$ and $K'$ is $C_{3v}$).  The  representation of the phonon modes at $\Gamma$ can be decomposed into $2E_1+E_2$, which corresponds to  two acoustic  modes (transforming according to one of the $E_1$ \emph{irreps}) and two optical modes transforming according to $E_1+E_2$. Hence, it  follows that the $K$ and $K'$ points contribute a total of $12$ optical modes, which transform according to $E'_1+E'_2+2G'$ in the (reducible) $C''_6$ vibrational representation.

\subsection{Central-force model for the in-plane phonons}\label{sec:centralforcemodel}

We next confirm and extend the previous findings based on
the symmetry analysis by explicitly obtaining the dispersion and wavefunctions for the phonons in the entire BZ. To this end, we use a simple model that assumes only
nearest-neighbor harmonic forces:~\cite{Guinea0,Castro}
\be
H_{ph}= \frac{1}{2}\sum_{i} m_i \dot{\boldsymbol{u}}^2_{i,\lambda} + \frac{1}{2} \sum_{ij} k_{ij} (\boldsymbol{\hat{\rho}}_{ij} \cdot(\boldsymbol{u}_i-\boldsymbol{u}_j))^2,\label{centralforces}
\ee
where atomic mass $m_i = m$ for the carbon atoms, $m_i=m_a$ for the adatoms (we take~\cite{Alicea1} $m_a > m$), $\boldsymbol{\hat{\rho}}_{ij}$ is the unit vector along the direction connecting  the $i$ and $j$ sites, $\boldsymbol{u}_i=(x_i,y_i)$ is the displacement of an atom at site $i$, and $k_{ij} =k$ is spring constant that determines the strength of the bond between two neighboring carbons. However, $k_{ij} =k'$ determines the force between neighboring adatoms and carbons. The above central-force model describes, for small atomic displacements,  the changes in kinetic and elastic energy.

\begin{figure}[t]
\centering
\includegraphics[width=8cm]{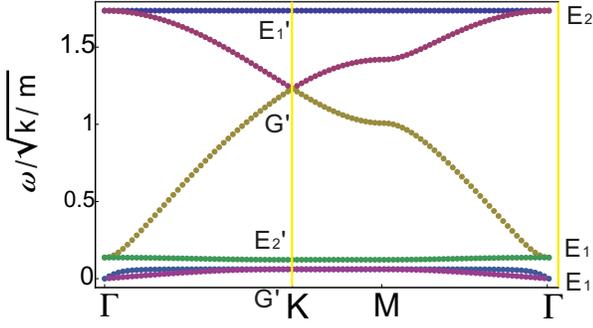}
\caption{Phonon spectrum along high symmetry points. When the unit cell is tripled, the $K$ and $K^{\prime}$ points are mapped to the $\Gamma$ point of the new BZ. The corresponding modes,  from lowest to highest energy, transfrom according to the \emph{irreps} $G', E_2', G'$ and $E_1'$, respectively. In addition,  the $\Gamma$ point of the original (three-atom) unit cell contributes with six modes (of the them acoustical) transforming according to the $E_1, E_1, E_2$ \emph{irreps}.}\label{fig2}
\end{figure}

 Let us next introduce a six-component vector of displacements, $\boldsymbol{u}(\boldsymbol{R})$, whose components are $\boldsymbol{u}(\boldsymbol{R})=\begin{bmatrix} x_A(\boldsymbol{R}) & y_A(\boldsymbol{R})  & x_B(\boldsymbol{R}) & y_B(\boldsymbol{R}) & x_0(\boldsymbol{R}) & y_0(\boldsymbol{R}) \end{bmatrix}^T,$ where $x_A, y_A, x_B$ and $ y_B$ correspond to the in-plane displacement   of the carbon atoms from their equilibrium positions on the A and B sublattices, respectively, and $x_0, y_0$ are displacements of adatom from its equilibrium H position; $\boldsymbol{R} = n_1  \boldsymbol{a}_1 + n_2 \boldsymbol{a}_2$ ($n_1,n_2$ being integers) spans the Bravais lattice (see Fig. \ref{fig_group}), whereas $\boldsymbol{\delta}_i$ describes the position of the atoms within the unit cell (i.e. the basis).  Note that we neglect out of plane dynamics, assuming that the system is on a substrate, which damps out the flexural phonons. Using periodic boundary conditions, the normal frequency $\omega_{\mu}$ in the three-atom unit cell can be computed by expanding $u_{i,\lambda}(\boldsymbol{R})$ in running waves ($\lambda=x,y$):
\be
{u}_{i,\lambda}(\boldsymbol{R})=\frac{1}{\sqrt{N}}\sum_{\mu=1}^6\sum_{\boldsymbol{q}} \sqrt{\frac{1}{m_i}}e^{ i \boldsymbol{q}\cdot (\boldsymbol{R}+\boldsymbol{\delta}_j)} \epsilon_{\mu,i,\lambda}( \boldsymbol{q}) u_{\mu}( \boldsymbol{q}),
\ee
where $\epsilon_{\mu,i,\lambda}( \boldsymbol{q})$ describes to the pattern of atomic displacements for the $\mu$th normal mode and satisfies $\boldsymbol{\epsilon}^{\dagger}_{\mu}( \boldsymbol{q})\cdot \boldsymbol{\epsilon}_{\mu'}(\boldsymbol{q}) =\delta_{\mu,\mu'}$ ($\boldsymbol{\epsilon}_{\mu}(\boldsymbol{q})$ is a six-component vector whose components
are $\epsilon_{\mu,i,\lambda}(\boldsymbol{q})$).  The explicit form of Eq.~\eqref{centralforces} in momentum space is given in the Appendix~\ref{a1}.   In terms of second quantization $u_{\mu}(\boldsymbol{q})=\sqrt{\hbar/2\omega_{\mu}}(b_{\mu}(\boldsymbol{q})+b_{\mu}^{\dagger}(-\boldsymbol{q})),$ the phonon Hamiltonian becomes
\be
H_{ph}=\sum_{\mu,\boldsymbol{q}}\hbar \omega_{\mu}(\boldsymbol{q})b_{\mu}^{\dagger}(\boldsymbol{q})b_{\mu}(\boldsymbol{q}).
\ee
The phonon spectrum along the high symmetry points is shown in Fig. \ref{fig2}. Two special symmetry points, $\Gamma$ and $K$(or $K'$), are marked by yellow lines and phonon modes at both points labelled
according to their irreducible representations.

 Let us next discuss the properties of the lowest energy phonon modes. The acoustic (Goldstone) modes whose energy vanishes at $\Gamma$ enter the  intra-valley electron-phonon coupling only through the space and time derivative of the mode displacement.~\cite{Basko} This is   because an uniform translation of the crystal cannot affect the motion of electrons (neglecting the coupling to the substrate).
On the other hand, the lowest-energy optical phonon at $\Gamma$ transforming according to the $E_1$ \emph{irrep} can  give rise to an electron-phonon coupling that is linear in the displacement field $\boldsymbol{u}(\boldsymbol{R})$. The square of the frequency for this mode is  given by the expression:
\be
\omega_{\Gamma}^2=\frac{3k'(2 m +m_a)}{2 m m_a}.\label{omegagamma}
\ee
Note that, for a heavy adatom (i.e. $m_a \gg m$), $\omega^2_{\Gamma} \to  3 k^{\prime}/2m$, which only depends on the coupling between the adatom and the carbon atoms, $k^{\prime}$,  and the carbon  atom mass, $m$. Typically, heavy ad-atoms are physisorpbed and therefore  $k^{\prime}\ll k$, meaning that the $E_1$ is expected
to have a rather low-energy, as we have indeed assumed
in the calculation shown in Fig.~\ref{fig2}. The polarization
vectors for the $E_1$ mode read:
\begin{small}
      \begin{align}
  \boldsymbol{\epsilon}_{I}(\Gamma) &=\frac{1}{\sqrt{m+m_a/2}}\left(\begin{array}{c}  -\frac{\sqrt{m_a}}{2}\\ 0\\ -\frac{\sqrt{m_a}}{2}\\ 0\\ \sqrt{m}\\ 0 \end{array} \right),\\
   \boldsymbol{\epsilon}_{II}(\Gamma)&=\frac{1}{\sqrt{m+m_a/2}}\left(\begin{array}{c} 0\\-\frac{\sqrt{m_a}}{2}\\ 0\\ -\frac{\sqrt{m_a}}{2}\\ 0\\ \sqrt{m} \end{array} \right).
    \end{align}
\end{small}

\begin{figure}[t]
\centering
\includegraphics[width=8cm]{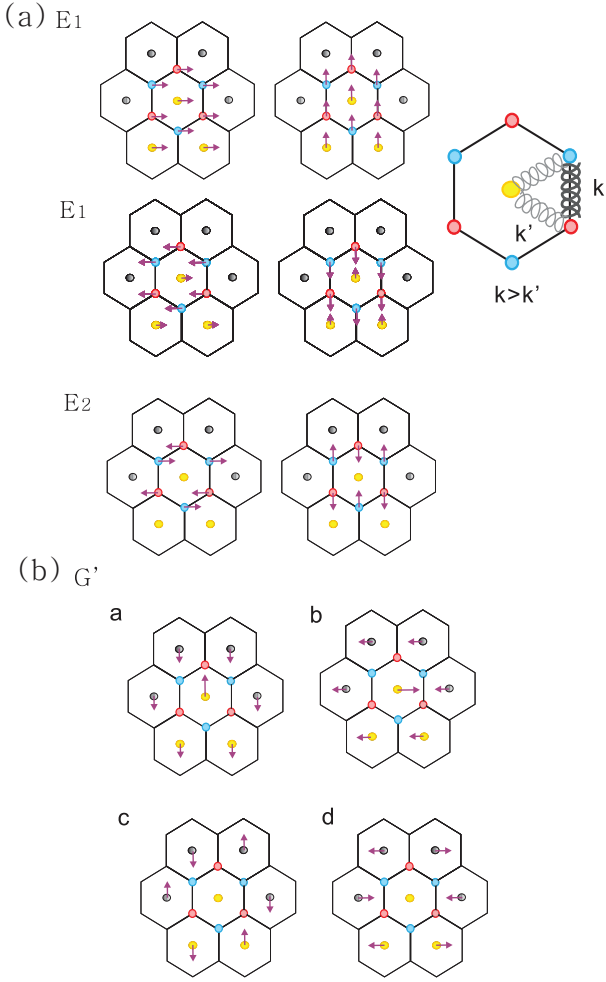}
\caption{(a)  Phonon modes $2E1+E_2$ at $\Gamma$ and (b) Phonon modes $G'=E_1+E_2$ at $K$ and $K'$. In (b) the small displacements of carbon atoms are not shown.}
\label{displacement}
\end{figure}

For crystal momentum $\boldsymbol{q}$ near $K (K')$,  phonons can scatter electrons from one Dirac point to the other.  The low-lying degenerate optical phonons transforming according to the  $G'$ \emph{irrep} have squared frequencies given by:
\be
\omega_{K (K^{\prime})}^2=\frac{3  k'}{2  m}+\frac{3 ( -b - \sqrt {b^2 - 4 a c }+2 k m_a )}{4  m_a m}\label{omegak}
\ee
The corresponding polarization vectors are:
\begin{align}
\boldsymbol{\epsilon}_1(K')&=B \left(\begin{array}{c} i\\ 1\\ -i\\ 1\\ 0\\ \frac{4 a}{-b+\sqrt{b^2-4ac}} \end{array} \right),\\
\boldsymbol{\epsilon}_2(K')&=B \left( \begin{array}{c} -1\\ i\\-1\\-i\\ \frac{4 a}{-b+\sqrt{b^2-4ac}}\\0 \end{array}\right)
 \end{align}
  where $B$ is the normalization constant, $b=-2 k' m + (k  + k') m_a, a=k'\sqrt{m_a m}$, and $c=-k'\sqrt{m_a m}$. The polarization vectors at $K$ are obtained from time-reversal symmetry which requires that $\epsilon_1(K')=\epsilon^*_1(K)$,  $\epsilon_2(K')=\epsilon^*_2(K)$.  Note that, for  $k\gg k^{\prime}$ and  $m_a\gg m$, $\epsilon_1(K')\to  [0,0,0 ,0,0, 1]^T$ and $\epsilon_2(K')\to  [0,0,0 ,0,1, 0]^T.$ That is, in the limit of weak adatom-carbon coupling and large adatom-carbon mass ratio, the mode is dominated by the motion
of the adatoms whereas the displacement of the carbon atoms becomes negligibly small (cf. Fig~\ref{displacement}b)

 When considering the modes at $K,K'$ points it is convenient to take their real and imaginary part combinations. The pattern of $E_1$ mode of $\Gamma$ point and $G'$ mode of $K$ and $K'$ are shown in Fig.~\ref{displacement}.
For Indium on Graphene, the value of adatom-carbon elastic constant has been estimated to be~\cite{private}
$k' \approx 0.02 \, \mathrm{Ry}/\mathrm{bohr}^2 \sim 1 eV/{\mathrm{\AA}}^2$ between carbon and Indium atoms with $k\approx 50$ eV$/{\mathrm{\AA}}^2$ between neighbor carbon atoms.
Hence, we obtain $\hbar \omega_{\Gamma}\simeq25$ meV and $\hbar \omega_{K}\simeq 11$ meV by Eq.~\eqref{omegagamma} and \eqref{omegak}, which is an energy scale comparable to the calculated~\cite{Alicea1} band gap $M=42$ meV for this system.
\section{Electronic structure of decorated graphene}\label{es}
A minimal tight-binding model for a single-layer graphene sheet uniformly covered by adatoms was given in Ref.~[\onlinecite{Alicea1}]. In what follows, we briefly review the main results obtained in Ref.~[\onlinecite{Alicea1}] that are relevant to our derivation of the phonon-SO coupling from this microscopic model. This will also allows us to introduce the continuum-limit Hamiltonian that describes the electronic states at low energies.

The Hamiltonian of the model contains three terms:
\be\label{eqHe}
H_0=H_g+H_a+H_c.
\ee
$H_g$ describes the hopping the  between the $\pi$ orbitals of the two nearest neighbor carbon atoms:
 \be
 H_{g}&=&-t\sum_{\langle \boldsymbol{r}\boldsymbol{r}'\rangle}(c^{\dagger}_{\boldsymbol{r}} c_{\boldsymbol{r}'} + \mathrm{h.c.}).\label{tightbinding}
  \ee
 The term allowing electrons to tunnel between the adatoms and its neighboring carbons can be written as follows:~\cite{Alicea1}
  \be
  H_{c}=-\sum_{I}\sum_{m=0,\pm1}t_{|m|}C_m^{\dagger}(\boldsymbol{R}_I)d_m(\boldsymbol{R}_I) + \mathrm{h.c.},\label{c_a}
 \ee
where $C_{m,\alpha}^{\dagger}(\boldsymbol{R}_I)=\sum_{j=1}^6\frac{1}{\sqrt{6}}e^{i m\pi (j-1)/3}c_{{\boldsymbol{r}_j}\alpha}^{\dagger}(\boldsymbol{R}_I)$.
As to the hopping amplitudes, $t_0$ is real and $t_1=t_{-1}$ purely imaginary; the index $I$ labels the plaquettes where the  adatoms sit on the H position. Finally, the Hamiltonian describing the adatoms reads:
 \be
 H_a&=&\sum_I d^{\dagger}(\boldsymbol{R}_I)h_a d(\boldsymbol{R}_I),
 \ee
 where
 \be\label{eqHa}
 h_a=\begin{bmatrix}
 \epsilon_1+\Lambda_{SO} s^z  & \sqrt{2}\Lambda'_{SO} s^- & 0 \\
 \sqrt{2}\Lambda'_{SO} s^+  & \epsilon_0 & \sqrt{2}\Lambda'_{SO} s^- \\
 0 & \sqrt{2}\Lambda'_{SO} s^+ & \epsilon_1-\Lambda_{SO} s^z
 \end{bmatrix}.
 \ee
and $d^{\dagger}(\boldsymbol{R}_I)=\begin{bmatrix} d_{1}^{\dagger}(\boldsymbol{R}_I), & d_{0}^{\dagger}(\boldsymbol{R}_I), & d_{-1}^{\dagger}(\boldsymbol{R}_I)\end{bmatrix}.$ The parameters $\epsilon_m, \Lambda_{SO}$ and $\Lambda'_{SO}$ are associated to crystal
field effects and spin-orbit coupling in the adatom $p$-orbitals.\cite{Alicea1}

 After pertubatively eliminating the adatom orbitals,~\cite{Alicea1} the low-energy Hamiltonian in the vicinity of the two Dirac points near $\boldsymbol{K} =(4 \pi/3\sqrt{3}a_0,0)$ (where $a_0 = 1.42$ {\AA} is the distance between two neighboring carbon atoms and $\boldsymbol{K} = -\boldsymbol{K}^{\prime}$) reads:
\be
H_0(\boldsymbol{k}) = \hbar v (\tau_z\sigma_x k_x+\sigma_y k_y)+M \tau_z\sigma_z s_z, \label{eq:h0}
\ee
where the last term $\tau_z\sigma_z s_z$ is the so-called Kane-Mele (or intrinsic) spin-orbit coupling. The Fermi velocity is $v=3t a_0 /(2\hbar)$ and the band gap $M =3\sqrt{3}\lambda_{so}$ (being $t = 2.7$ eV in single-layer graphene). A linear Rashba correction to $H_0(\boldsymbol{k})$ of the form
\begin{align}
H_{cr} &= -\beta_{R}(k_xs_y-k_y s_x)+\beta_{R} \left[k_x (s_x \sigma_y\tau_z - s_y\sigma_x)  \right. \nonumber \\
& \left. - k_y (s_x \sigma_x + s_y\sigma_y \tau_z)\right]+ \cdots\label{eq:hr}
\end{align}
 with $\beta_{R}=9\lambda_R a_0 /2$ is also induced by the adatoms. However, this correction is not important near the Dirac points at $K$ and
 $K^{\prime}$ and can dropped because it yields subdominant corrections to the band structure, which otherwise realizes the Kane-Mele model.~\cite{Alicea1} For Indium coverage,  $\lambda_{so}\simeq 8$ meV and $\lambda_R=20$ meV, so $M=42$ meV and $\beta_R/a_0=90$ meV. The above continuum model provides an accurate description of the electronic structure for energies well below the cut-off $E_c = \hbar v/a_0\simeq 4.1$ eV.

\section{Phonon-SO coupling}\label{sec:coupling}
\subsection{Group Theoretic Analysis}

 In order to obtain the form of the phonon-SO interaction, the phonon displacement fields must be combined with electron spin and orbital
operators in order to yield a scalar under the symmetry group $C''_{6v}$. Furthermore, symmetry requires that the
phonon field and the  electron operators transform according
to the same \emph{irrep}. In this work, we are interested in the coupling of the $\pi$-electrons   near $K (K')$, which transform according to the
8-dimensional irreducible representation of $C''_{6v}$. We take three different sets of $2\times2$ Pauli matrices ${\sigma}, {\tau}$ and $\boldsymbol{s}$ to define the physical sublattice isospin, valley pseudospin, and real spin degrees of freedom respectively and these matrices act on eight-component  spinor:
\be
\Phi(\boldsymbol{k})=
\left(
\begin{array}{c}
 c_{ A,\boldsymbol{K},\uparrow}(\boldsymbol{k}) \\
   c_{ B,\boldsymbol{K},\uparrow}(\boldsymbol{k})\\
   c_{ A,\boldsymbol{K}',\uparrow}(\boldsymbol{k})\\
   c_{ B,\boldsymbol{K}',\uparrow}(\boldsymbol{k})\\
 c_{ A,\boldsymbol{K},\downarrow}(\boldsymbol{k})\\
    c_{ B,\boldsymbol{K},\downarrow}(\boldsymbol{k})\\
   c_{ A,\boldsymbol{K}',\downarrow}(\boldsymbol{k})\\
    c_{ B,\boldsymbol{K}',\downarrow}(\boldsymbol{k})
 \end{array} \right)\label{spinorab}
\ee
   The possible SO operators transforming according to \emph{irreps} of $C''_{6v}$ that are even under time-reversal symmetry are displayed in table~\ref{c6pauli}.~\cite{Basko,Edward,ochoa} Note that the operators that contain $s_z$ are even under mirror symmetry operation, $z\rightarrow-z$ with respect to the graphene plane. On the other hand, the operators containing $s_x$ or $s_y$ are odd. The different forms of these operators for different choices of the spinor basis are discussed in Refs.~\onlinecite{Basko,Edward,ochoa},
together with the details of the derivation of the table results.
\begin{table}[t]
\centering
\begin{tabular}{|c|c|p{110pt}|}\hline
\emph{Irrep}& $z\rightarrow -z$ even &  \qquad $z\rightarrow -z$  odd \\\hline
$A_1$& $\sigma_z \otimes \tau_z \otimes s_z $& $\sigma_x \otimes \tau_z \otimes s_y-\sigma_y \otimes s_x $\\\hline
$A_2$&  &  $\sigma_x \otimes \tau_z \otimes s_x+\sigma_y \otimes s_y $   \\\hline
$B_2$& $ \tau_z \otimes s_z$  &  \\\hline
$E_1$& $\begin{pmatrix}
-\sigma_y \otimes  s_z \\\sigma_x \otimes \tau_z \otimes s_z \end{pmatrix}$
& $\begin{pmatrix}
-\sigma_z \otimes \tau_z \otimes s_y \\ \sigma_z \otimes \tau_z \otimes s_x
\end{pmatrix}$ \\\hline
$E_2$&  &  $\begin{pmatrix}
\sigma_x \otimes \tau_z \otimes s_y+\sigma_y  \otimes s_x   \\\sigma_x \otimes \tau_z \otimes s_x-\sigma_y  \otimes s_y
\end{pmatrix}$
$ \begin{pmatrix}
-\tau_z \otimes s_y \\\tau_z \otimes s_x
\end{pmatrix} $ \\\hline
$E_1'$& $\begin{pmatrix}
-\sigma_y \otimes \tau_y \otimes s_z \\
\sigma_y \otimes \tau_x \otimes s_z
\end{pmatrix}
 $ & \\\hline
$G'$&   & $\begin{pmatrix}
-\sigma_y \otimes \tau_y \otimes s_x \\
-\sigma_y \otimes \tau_y \otimes s_y \\
\sigma_y \otimes \tau_x \otimes s_y \\
\sigma_y \otimes \tau_x \otimes s_x
\end{pmatrix}$  \\\hline
\end{tabular}
\caption{Classification of possible SO coupling terms under the $C''_{6v}.$ All these terms preserve time-reversal symmetry}\label{c6pauli}
\end{table}

The phonon SO coupling can be expanded in powers of the phonon displacement fields and its derivatives. In what follows, we shall focus in the long wavelength limit, that is, we assume that the momentum transferred by the phonon  ($\boldsymbol{q}$) is small (i.e. $|\boldsymbol{q}|\ll a^{-1}_0$).  In what follows, we consider only interactions involving one phonon and derivative couplings will be neglected. In this regard, we recall that, at the $\Gamma$ point, a uniform translation of the acoustic ($E_1$) phono does not affect the electron dynamics and therefore, the coupling to these phonon modes is necessarily derivative and can be neglected for $|\boldsymbol{q}|\to 0$. In addition, we note that there is no term in Table~\ref{c6pauli} which transforms according to the $E_2'$ and respects the time reversal symmetry simultaneously. Therefore,  the leading term from $E_2'$ must be dependent quadratically on the atomic displacement and it will be neglected in out treatment.

Using the above results, we can classify the electron-spin to phonon coupling into two types: inter- and intra-valley couplings. By combining  the phonon displacement field operators with the SO operators from table~\ref{c6pauli} that transform according to the same \emph{irrep}, we can obtain the leading-order couplings between the phonons to the electron spin. Thus, the inter-valley  coupling takes the form:
\be
H_{pSO, K}&=&A_K\sigma_y (- u_{K,a} s_y  \tau_y -
   u_{K,b} s_x  \tau_y\nonumber\\ &+&u_{K,c} s_y  \tau_x+ u_{K,d} s_x  \tau_x ).\quad \label{eq:inter}
\ee
Whereas the intra-valley coupling reads:
\begin{small}
\be
H_{pSO,\Gamma}&=&C_{\Gamma } (u_{\Gamma, I} s_x-u_{\Gamma, II} s_y) \sigma_z \tau_z\nonumber\\ &+& D_{\Gamma }(+u_{\Gamma, I} \sigma_y-u_{\Gamma,II} \sigma_x \tau_z)s_z .\label{eq:intra}
\ee
\end{small}
 In the above expressions, we have restricted our attention to the lowest frequency optical mode transforming according to the $G'$ \emph{irrep} from $K (K')$ and the $E_1$ mode from $\Gamma$, ignoring  higher frequency modes (such like the $E^{\prime}_1$ or the other $G'$ mode). Although the coupling of between electron-spin and those high-frequency modes is not expected to vanish, their relatively high energy yields an scattering rate with the electrons that is (more strongly) exponentially suppressed at low temperatures. In other words, those phonon modes are less important in the low-energy sector that concerns us here and, therefore, we shall not discuss them any further.

 As to the optical mode transforming according to $E_2'$ near the $K (K^{\prime})$ point, even though its frequency is small compared to other optical modes, we shall not consider it.  The reason is that the lowest order coupling between the electron-spin and this mode does not preserve time reversal symmetry. This can be seen from table~\ref{c6pauli}: It is not possible to write down
a time-reversal invariant operator involving the electron spin
that transforms according to the $E^{\prime}_2$ \emph{irrep}.

\subsection{Microscopic derivation}\label{sec:md}

In what follows we shall extend the tight binding model introduced in Sect.~\ref{es}  beyond the rigid-lattice (Born-Oppenheimer) approximation. This will allow for an alternative (re-)derivation of the form of the electron-spin to phonon couplings. In addition, we shall be able to estimate the strength of couplings of  Eqs.~(\ref{eq:inter},\ref{eq:intra}), which is not possible from the symmetry analysis provided in the previous section.

 In order to obtain the electron-phonon coupling, we need to expand the hopping amplitudes in Eq.~\eqref{eqHe} as a function of the atomic positions. Thus, we obtain two types of electron-phonon couplings. The first type results from the modulation in length of the carbon-carbon bonds and is also present in pristine graphene. It can be written as follows:~\cite{Castro,Katsnelson}
\be
H^{ph}_{t}&=&-\sum_{\langle \boldsymbol{r},\boldsymbol{r}'\rangle}\left\{
\frac{\partial t}{\partial{l}}\hat{\boldsymbol{b}}_{\boldsymbol{r}\boldsymbol{r}'}\cdot(\boldsymbol{u}_{\boldsymbol{r}}-\boldsymbol{u}_{\boldsymbol{r}'})c^{\dagger}_{\boldsymbol{r}} c_{\boldsymbol{r}'}+ \mathrm{h.c.}\right\},\qquad\label{phononc_c}
\ee
where $l$ is the interatomic distance and $\hat{b}_{\boldsymbol{r}_1\boldsymbol{r}_2}=\hat{b}_{\boldsymbol{r}_5\boldsymbol{r}_4}= (-\frac{\sqrt{3}}{2},-\frac{1}{2})a_0$, $\hat{b}_{\boldsymbol{r}_3\boldsymbol{r}_2}=\hat{b}_{\boldsymbol{r}_5\boldsymbol{r}_6}= (0,1) a_0$ and $\hat{\boldsymbol{b}}_{\boldsymbol{r}_3\boldsymbol{r}_4}=\hat{b}_{\boldsymbol{r}_1\boldsymbol{r}_6}=(\frac{\sqrt{3}}{2},-\frac{1}{2}) a_0$ are the vectors connecting a carbon  atom on the B sublattice to its nearest neighbors on the A. The second type stems from the modulation of distance between the adatoms and their neighboring carbons, and
it is described by a term of the form:
 \begin{small}
 \be
  H^{ph}_{c}=-\sum_{I}\sum_{m=-1}^{+1}  \frac{\partial t_{|m|}}{\partial{l}} \hat{\delta}_{\boldsymbol{r}_j}\cdot\Delta\boldsymbol{u}_{\boldsymbol{r}_j} C_{m}^{\dagger}(\boldsymbol{R}_I)d_m(\boldsymbol{R}_I)+
  \mathrm{h.c.},\quad \label{phononc_a}\ee
 \end{small}
where $\delta_{\boldsymbol{r}_1}=-\delta_{\boldsymbol{r}_4}= (0,-1) a_0,\delta_{\boldsymbol{r}_3}=-\delta_{\boldsymbol{r}_6}= (\frac{\sqrt{3}}{2},\frac{1}{2}) a_0$ and $\delta_{\boldsymbol{r}_5}=-\delta_{\boldsymbol{r}_2}=(-\frac{\sqrt{3}}{2},\frac{1}{2}) a_0$ are the vectors connecting the adatom to its neighboring carbons, and $\Delta\boldsymbol{u}_{\boldsymbol{r}_j}=\boldsymbol{u}_{\boldsymbol{r}_j}-\boldsymbol{u}_0$ with $\boldsymbol{u}_{\boldsymbol{r}_j}$ and $\boldsymbol{u}_0$ associated to the displacement operator of carbons and adatoms respectively. Both types of electron-phonon couplings contribute to the  electron-phonon scattering rate but only the latter one gives rise to the spin-phonon coupling in which we are interested here.

In order to study the coupling within the low energy sector, we can integrate out the adatom's degrees of freedom
by performing an unitary transformation~\cite{Alicea1,Yao} (see Appendix~{\ref{a2}). Hence, the strength of the phonon to electron-spin coupling can be obtained to be~(see Eqs.~\eqref{eq:inter} and \eqref{eq:intra})
\be
 C_{\Gamma }&=&\frac{3(2+\alpha^2)\Lambda'_{so}|t_1 \frac{\partial t_{0}}{\partial l} |}{2\sqrt{2}\sqrt{ m+m_a/2} \alpha \epsilon_0\epsilon_1}\label{coupling_C},\\
D_{\Gamma}&=&\frac{3(2+\alpha^2)\Lambda_{so}|t_1 \frac{\partial t_{1}}{\partial l} |}{4 \sqrt{m+m_a/2}\alpha \epsilon_1^2}\label{coupling_D},
\ee
and
\begin{small}
\be
A_K=\frac{(3 \alpha b'-3 \alpha\sqrt{b'^2+4}+6) \Lambda'_{so}|t_1 \frac{\partial t_{0}}{\partial l} |}{\sqrt{2 m_a }\epsilon_0\epsilon_1 (b'-\sqrt{b'^2+4})\sqrt{ 4+b'^2+b'\sqrt{b'^2+4}}},\label{coupling_A}
\ee
\end{small}
where $b'=b/a=-
\frac{2 }{\alpha} + \frac{(k  + k') \alpha}{k'} $ with $\alpha=\sqrt{m_a/m}.$

\textcolor{black}{
Let us next estimate numerically the magnitude of these couplings.
In pristine graphene, $\beta=\partial \ln t/ \partial \ln l\simeq 2$ to $3$.~\cite{Katsnelson} Thus,  we choose $\beta=2$ and estimate $\partial  t_0/ \partial  l \approx \beta t_0/a_0$ and $\partial  t_1/ \partial  l \approx \beta t_1/a_0$. Using the parameters given in Ref.~[\onlinecite{Alicea1}], we estimate the intra-valley $s_z$-conserving electron-phonon coupling  to be $D_{\Gamma} \approx 46$ meV/{\AA} $\sqrt{m}$, intra-valley spin-flip electron-phonon coupling  $C_{\Gamma} \approx 98$ meV /{\AA}$\sqrt{m}$,
and the inter-valley spin-flip coupling $ A_{K} \approx 28$ meV/{\AA}$\sqrt{m}$. In addition, using Eq.~\eqref{eq:inter} and Eq.~\eqref{eq:intra}, we can study the effective electron-electron interaction by integrating out the phonon modes $u_{\Gamma, I}, u_{\Gamma, II}, u_a,u_b,u_c,u_d$. The dominative effective interactions induced by phonon are: $C_{\Gamma}^2/2\omega^2_{\Gamma}\sim 2.6$ meV $D_{\Gamma}^2/2\omega^2_{\Gamma} \sim 0.57$ meV  and $A_K^2/2\omega^2_K\sim 1.3$ meV.}
%
\section{Application: Quasi-particle Spin-flip Rate due to Phonons}
\label{sec:spinres}

\subsection{Formalism}

The electron-spin to phonon coupling displayed in Eqs. (\ref{eq:inter}) and (\ref{eq:intra}) is the main result of this work. It can be applied to different problems,  such like the calculation of spin transport in the system as well as the study of superconducting pairing instabilities in ad-atom decorated graphene. However, in this section, we focus on computing the spin-flip rate of electrons and holes. This will provide a quantitative estimation of the importance of this coupling as a function of the quasi-particle energy and system temperature.

Indeed, the existence of the spin-electron phonon coupling implies that phonon emission and absorption can flip the spin of electronic quasiparticles. The spin-flip rate, $\Gamma$, can be obtained from the imaginary part of the electron self-energy~($\Sigma$) as $\Gamma = -\mathrm{Im} \: \Sigma$. The self-energy
 is related to the single-particle Green's function ($G$) by means of Dyson's equation: $G^{-1}=G_0^{-1}-\Sigma$, where $G_0$ is the non-interacting electron Green's function.  The  quasiparticle decay rate, $\Gamma$, is a sum of different contributions $\Gamma=\Gamma_{\mathrm{ee}}+\Gamma_{\mathrm{dis}}+\Gamma_{\mathrm{e-ph}}+\cdots$, which  contains contributions from electron-electron interactions $\Gamma_{ee}$, disorder  $\Gamma_{\mathrm{dis}}$,  lattice vibrations $\Gamma_{e-ph}$, and other inelastic mechanisms.
Here we focus on $\Gamma_{\mathrm{e-ph}}$ term, which contains contributions from  two different channels: the spin-flip channel~($\Gamma^{\mathrm{flip}}$) and non-spin-flip channel~($\Gamma^{\mathrm{non}}$). The non-spin-flip channel contribution arises from the electron-phonon couplings of various kinds. On the other hand,  the spin-flip contribution is exclusively generated by the spin-phonon coupling that has been discussed above. Note that such phonon-mediated spin-flip processes are different from those due to electron-electron interactions and (magnetic/SOC) impurity scattering because they typically have a much more pronounced temperature and energy dependence due to the optical nature of the phonons involved. Therefore, the understanding of this spin-flip processes becomes necessary when thinking about possible device applications, which must be operated in a finite temperature.

In order to investigate the spin-flip rate due to phonons, we first compute the (non-interacting) quasi-particle dispersion by  diagonalizing the Hamiltonian in Eq.~\eqref{eq:h0}. This can be achieved  by means of a unitary transformation, $U$~(see Appendix~\ref{a3} for the details):
\begin{align}
\Psi(\boldsymbol{k}) = U^{-1} \Phi(\boldsymbol{k})=
\left(
\begin{array}{c}
c_{ +,\boldsymbol{K}',\downarrow}(\boldsymbol{k})\\
    c_{ +,\boldsymbol{K},\downarrow}(\boldsymbol{k})\\
   c_{ +,\boldsymbol{K}',\uparrow}(\boldsymbol{k})\\
   c_{ +,\boldsymbol{K},\uparrow}(\boldsymbol{k})\\
  c_{ -,\boldsymbol{K}',\downarrow}(\boldsymbol{k})\\
    c_{ -,\boldsymbol{K},\downarrow}(\boldsymbol{k})\\
   c_{ -,\boldsymbol{K}',\uparrow}(\boldsymbol{k})\\
    c_{ -,\boldsymbol{K},\uparrow}(\boldsymbol{k})
  \end{array}
    \right),
 \end{align}
where $\Phi(\boldsymbol{k})$ is the original basis given by Eq.~\eqref{spinorab}. As a result, the unperturbed Hamiltonian  becomes diagonal, i.e.
\be
U^{\dagger}H_0(\boldsymbol{k})U-\mu \hat{N}=\begin{bmatrix}  \varepsilon_{\boldsymbol{k},+} I_{4\times 4} &  0 \\
 0 &  \varepsilon_{\boldsymbol{k},-} I_{4\times 4} \end{bmatrix}\hat{N}\label{foursate}
 \ee
where $\varepsilon_{\boldsymbol{k},\pm}=\pm\sqrt{M^2+\hbar^2 v^2 \boldsymbol{k}^2}-\mu$. Here $\mu$ is the chemical potential and $I_{4\times 4}$ is the $4\times 4$ unit matrix.

Since the electron spin is coupled to the  phonon by  means of \eqref{eq:inter} and \eqref{eq:intra}, which are both linear in atomic displacement~(i.e. phonon field), the decay rate including spin-flip and non spin-flip processes can be obtained by calculating the imaginary part of the electron self-energy to lowest order in the electron-phonon interaction~\cite{coleman}. We shall neglect the renormalization of the quasi-particle energy, which effectively renders our approach entirely equivalent to a Fermi-Golden rule calculation. Thus, the decay rate for a quasi-particle~(or quasi-hole) with  energy $\omega$~(referred to Fermi energy) and band-index $\pm$~(where $+$ stands for conduction and $-$ for valence band), valley $K(K')$ and spin $\uparrow/\downarrow$ at momentum ${\bf k}$ is given by
\be
\Gamma_{\boldsymbol{k},\eta}(\omega)&=&
 \frac{2 \pi}{\hbar}\sum_{\boldsymbol{p}, \eta_1}\sum_{\alpha}\frac{ J_{\alpha}^2}{2\omega_{\alpha} N} (U^{\dagger}{L}^{\alpha}U)_{\eta,\eta_1}
(U^{\dagger}{L}^{\alpha}U)_{\eta_1,\eta}  \times \nonumber\\
&&\Big[\big(1+n(\omega_{\alpha})
-f(\varepsilon_{\boldsymbol{k}-\boldsymbol{p},\eta_1})\big) \delta\big(\omega-(\varepsilon_{\boldsymbol{k}-\boldsymbol{p},\eta_1}+\omega_{\alpha})\big)
\nonumber\\&&+\big(n(\omega_{\alpha})+f(\varepsilon_{\boldsymbol{k}-\boldsymbol{p},\eta_1})\big)\delta\big(\omega-(\varepsilon_{\boldsymbol{k}-\boldsymbol{p},\eta_1}-\omega_{\alpha})\big)\Big].
\nonumber\\\label{eq:FermiGorden}
\ee
where $\eta=(\pm,K (K'), \uparrow(\downarrow))$ denotes the set of quasi-particle quantum numbers (band, valley, spin), and $N$ is the total number of unit cells. $L^{\alpha}$ stands for the $4\times4$ matrices in Eq.~\eqref{eq:inter} and the two terms in Eq.~\eqref{eq:intra} whose couplings are $J_{\alpha}=A_K, C_{\Gamma},D_{\Gamma}$. The phonon frequencies $\omega_{\alpha}=\omega_{\Gamma},\omega_K$ are given in Eqs.~(\ref{omegagamma},\ref{omegak}); $f(\omega)$ and $n(\omega)$ are the Fermi-Dirac and Bose-Einstein distributions, respectively.


\begin{figure}[htbp!]
\centering
\includegraphics[width=8.5cm]{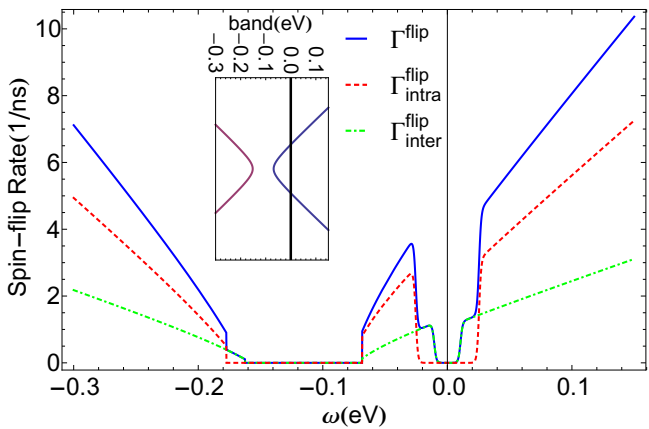}
\caption{Quasi-particle spin-flip rate due to the electron spin-phonon coupling, as a function of quasi-particle energy $\omega$  when $T=10K,$  the band gap is $\Delta = 0.1$ eV and chemical potential around $0.11$~eV, measured from the middle of the spin gap.  The blue line is for the total spin-flip rate $\Gamma^{\mathrm{flip}}(\omega)=\Gamma^{\mathrm{flip}}_{\mathrm{intra}}(\omega)+\Gamma^{\mathrm{flip}}_{\mathrm{inter}}(\omega)$,  red dashed line for intra-valley spin-flip $\Gamma^{\mathrm{flip}}_{\mathrm{intra}}(\omega)$  and green dash-dotted line for intervalley spin-flip $\Gamma^{\mathrm{flip}}_{\mathrm{inter}}(\omega).$    The inset shows the band structure  near a Dirac point with a spin gap with a black thick line associated with the location of the chemical potential. For this plot, we use the parameters of Indium coverage discussed in context. Note that $\Gamma^{\mathrm{flip}}_{\lambda}(\omega), \Gamma^{\mathrm{flip}}_{\mathrm{intra},\lambda}(\omega), \Gamma^{\mathrm{flip}}_{\mathrm{inter},\lambda}(\omega)$ can have values only when $\lambda(\omega+\mu)\geq M$, but for convenience we plot them together to show overall features among the whole range of $\omega.$}\label{fig4}
\end{figure}

\begin{figure}[htbp!]
\centering
\includegraphics[width=8cm]{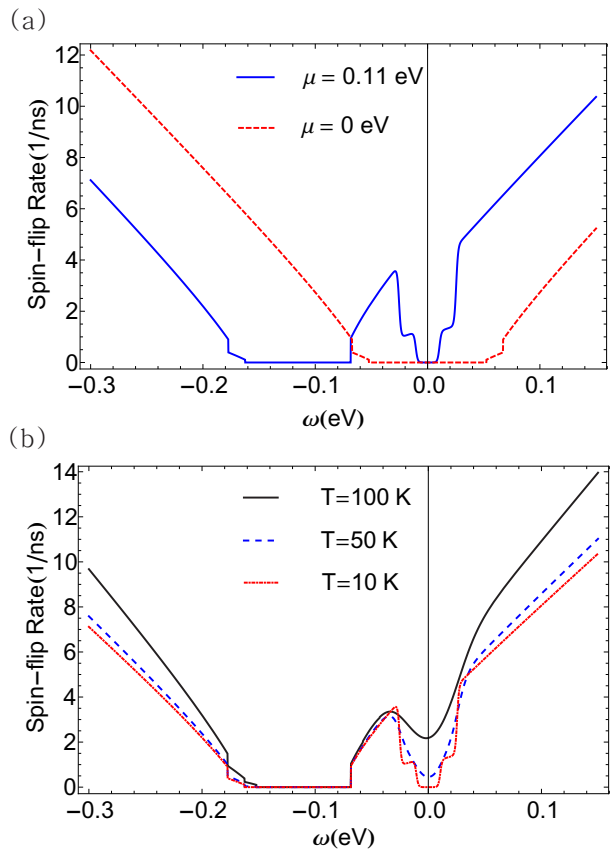}
\caption{ Spin-flip rate $\Gamma^{\mathrm{flip}}(\omega)$ as a function of the quasi-particle energy $\omega.$ (a) for chemical potentials $\mu=0.11$~eV and $0$ eV for $T=10K.$ (b) for various temperature $T=10, 50, 100K$  for $\mu=0.11$~eV. The values of the remaining parameters are  the same as in Fig.~\ref{fig4} }\label{fig5}
\end{figure}

\subsection{Quasi-particle Spin-flip rate}

 The spin-flip rate of a spin polarized quasi-particle, say, an electron injected at any given valley with spin up and energy $\hbar\omega$~(relative to the Fermi energy),  may be easier to determine experimentally.
Note that, as mentioned above, we are interested in the contribution from the spin-flip rate as obtained from the electron spin-phonon coupling derived above~(processes that do not flip the spin can also be induced by other mechanisms such like the normal electron-phonon coupling). Thus, we shall retain the contribution from the following four matrix elements:
 $(U^{\dagger}{L}^{\alpha}U)_{(a,\uparrow),(b,\downarrow)}$ in Eq. \eqref{eq:FermiGorden}, where $a,b=(\pm,K/ K')$.

For the current example, the band structures are isotropic for momentum $\boldsymbol{k}$ and symmetric between unlike spins at low energy. Therefore,  after some manipulations~(see  Appendix~\ref{a4} for details), the spin-flip rate can be shown to be independent on the direction of momentum $\boldsymbol{k}$ and the spin of the quasi-particle.  As a consequence, the spin-flip rate of a spin polarized quasi-particle around Fermi surface can be written as
\be
\Gamma^{\mathrm{flip}}_{\lambda}(\omega)&=&\Gamma^{\mathrm{flip}}_{\mathrm{inter},\lambda}(\omega)+\Gamma^{\mathrm{flip}}_{\mathrm{intra},\lambda}(\omega)\nonumber\\
&=&\frac{2 \pi}{\hbar}\left[2{ A_K^2}L_{\lambda}(\omega,\omega_K)+{ C_{\Gamma}^2}L_{\lambda}(\omega,\omega_{\Gamma})\right]\label{eq:spindecayrate},
\ee
where $\lambda=\pm$ refers to the band index.
 $\Gamma^{\mathrm{flip}}_{\mathrm{inter},\lambda}$ and $\Gamma^{\mathrm{flip}}_{\mathrm{intra},\lambda}$ are the contributions of the inter-valley and intra-valley phonon SO couplings~(see Eqs. (\ref{eq:inter}) and (\ref{eq:intra})) respectively. Here $L_{\lambda}(\omega,\omega_{\alpha})$ is defined by~(see Appendix D)
\begin{align}
&L_{\lambda}(\omega,\omega_{\alpha})=\Theta(\lambda \hbar \omega'-M)\frac{\Omega}{ 2 N  \hbar\omega_{\alpha} \pi v^2 } \notag\\
& \times \left\{
\Big[1+n(\omega_{\alpha})-f(\omega-\omega_{\alpha})\Big]
\Big[ \hbar\big|\omega'-\omega_{\alpha}\big|-\frac{ M^2}{  \hbar |\omega' |}\Big] \right.
\notag\\
&\qquad \times \Theta (\hbar^2(\omega'-\omega_{\alpha})^2-M^2) \notag\\
&\quad   + \Big[n(\hbar\omega_{\alpha})+f(\omega+\omega_{\alpha})\Big]\Big[ \hbar \big|\omega'+\omega_{\alpha}\big|-\frac{ M^2}{  \hbar |\omega' |}\Big] \notag\\
&\left.  \quad \times \Theta(\hbar^2(\omega'+\omega_{\alpha})^2-M^2)\right\},
\label{eq:g}
\end{align}
where $\Omega$ is the system size (area) and $N$ is the number of unit cells.  $\Theta(x)$ is Heaviside step function, and $\omega'\equiv\omega+\mu$.
The ratio between coefficients in front of $A_K^2$ and $C_{\Gamma}^2$ in Eq. \eqref{eq:spindecayrate} result from the fact that Eq.~(\ref{eq:inter}) contains four  terms proportional to $s_x, s_y$, which can cause spin-flip, whereas Eq.~(\ref{eq:intra}) contains only two such terms.

In Fig.~\ref{fig4},  we have plotted the quasi-particle spin-flip rate as a function of the  quasi-particle energy, $\omega$ ~(relative to the Fermi energy). The Fermi energy (chemical potential at $T = 0$), $\mu$, is measured from the middle of the band gap. We have set the temperature $T=10$~K. The band structure  near a Dirac point is shown in the inset. The quasiparticle spin-flip rate vanishes between $\omega \simeq -0.16$~eV and $\omega \simeq -0.07$ eV because quasiparticle propagation is not possible within the band gap. Furthermore, energy conservation suppresses  the phonon emission at low temperatures: In order to emit a phonon, the electron quasi-particle energy~($\omega>0$) must satisfy the condition that $\omega=\varepsilon_{\boldsymbol{k}} =\varepsilon_{\boldsymbol{k}-\boldsymbol{q}}+\omega_{K (\Gamma)}$. For  a quasi-hole~(i.e. for $\omega<0$)  we must have  $\omega=-\varepsilon_{\boldsymbol{k}}=-\varepsilon_{\boldsymbol{k}-\boldsymbol{q}}+\omega_{K/\Gamma}$. In addition, Fermi statistics further prevents an electron from emitting a phonon if the final state is below the Fermi energy. As a consequence, since the electron spin-flip is caused by optical phonons with energies~$ \omega_{K}$ and $\omega_{\Gamma} > \omega_{K}$ (for $q \to 0$),  the quasi-particle spin-flip rate exhibits a two-step variation close to $\omega=0$ in Fig.~\ref{fig4}. The phenomena applies to holes: Because of Fermi statistics, a double step variation occurs only near  the top of the valance band and near (but below) the Fermi energy.  The variation of the spin-flip rate with the chemical potential, $\mu,$  is shown in In Fig.~\ref{fig5}(a) for $T = 10$ K. The temperature dependence, however, is displayed  in Fig.~\ref{fig5}(b) for  $\mu\approx 0.11 eV$. The double step structure still occurs for different values of $\mu$. However,  increasing the temperature  smears out this structure.

Finally, we briefly remark the effects of the Rashba coupling~(cf. Eq.~\eqref{eq:hr}), which has been neglected when calculating the spin flip rate due to the coupling with phonons. There are two different influences of this term on the spin motion: First, the presence of the Rashba term in the band structure leads corrections to the inelastic spin flip rate through the change of the density of states~(i.e. the slope of the energy dispersion at low energies). The effect can be estimated by replacing the prefactor $1/v^2$ in Eq.~\eqref{eq:g} by $1/(v\pm \beta_R/\hbar)^2\approx1/v^2\mp2\beta_R^2/\hbar^2 v^2$.   Thus we see that the correction arising from the linear Rashba term is  $\sim \beta_R^2/\hbar v^2 \approx 10^{-3}$, which can be safely neglected.

The second effect results from the fact that the electron spin is no longer a good quantum number when the Rashba term is included. In other words, an injected electron whose spin is pointing  up in the out of the plane direction, for instance, will naturally precess even {\it without} coupling to the phonons. In other words, the spin state will experience Rabi-like oscillations. Nevertheless, this Rashba-induced precession does not lead to energy loss of the injected electron and therefore, it should not affect the inelastic spin flip rate mediated by phonons that we have computed above. We can estimate the precession rate~($\Gamma_R$) by diagonalizing the electron Hamiltonian including the order correction to order $\beta_R /\hbar v$:
\begin{equation}
 \Gamma_R/\mu\approx  \frac{1}{\hbar}\frac{\beta_R}{\hbar v}\frac{\mu}{M},
\end{equation}
where $\mu$ is the chemical potential. If we use the same parameters as in Fig. \ref{fig4} taking $\mu\simeq 100$ meV, we find $\Gamma_R\sim 10^4$ ns$^{-1}$, which is much larger than the spin flip rate~($\Gamma_{\rm flip}$) due to spin-phonon interaction. Therefore, in an experiment, the spin orientation of the injected electron will precess very fast due to the spin-orbital (Rashba) coupling. However, such precession 
will be damped by inelastic processes arising from e.g. the coupling to the phonons. Thus, the double step structure as a function of the quasi-particle energy that we found above should still appear even in the presence of a Rashba-type spin-orbit coupling.

\section{Discussion}\label{sec:Discussion}

We have obtained the electron-spin to phonon coupling
for a single-layer of graphene uniformly covered with heavy adatoms. The
coupling is obtained using both symmetry arguments
and a simple tight-binding model. However, when
considering the experimental consequences of our calculations,
a major concern is that, as it has been
found (see e.g.~Ref.~[\onlinecite{kawakami,jaja2}]),
adatoms tend to cluster rather than uniformly
covering the graphene layer. Indeed, it has been discussed
by a number of authors recently, some of the features of the~(topological) band gap may be still observable in the quasi-particle spectrum~\cite{Paul} or in the transport properties~\cite{Niu,Roche}
provided clustering is not very strong~\cite{Roche}.
In such scenario,  we expect that, even in the presence of a sizable randomness in adatoms coverage, a coupling between the electron spin
and the lattice dynamics exists. Assuming that a large faction of the adatoms are located in the $H$ position, we can expect that the
randomness will lead to some sizable broadening of the phonon spectrum
(especially of the shorter wave-length phonons at the $K$ and $K^{\prime}$ points). The latter will result in a rounding up  of
the sharp features of the energy dependence of
spin-flip rate shown in Fig.~\ref{fig4} and \ref{fig5}, even
at zero temperature. For a clean and uniformly covered single-layer graphene system,  the existence of these sharp features and the exponential suppression of the spin-flip processes that they imply for quasi-particles near the Fermi level seem to imply that the edge transport may be essentially ballistic temperatures, if disorder exists in the adatom coverage, this may not be longer the case. The quantification of this effect is currently under investigation~\cite{unpub}.

In addition, our results also open the door to research into superconductivity in decorated graphene. Indeed, in recent time, studies of  layer materials such as graphene heterostructures separated by hexagonal boron
nitride~\cite{Guinea} and molybdenum disulfide~\cite{Roldan} were proposed, when the carrier concentration is high and the screening of long range Coulomb potential is strong, to have a superconducting phase, at which the cooper pairing has opposite signs in different valleys. It becomes more interesting in the graphene decorated with heavy adatoms because the couplings between the
electron-spin and lattice vibration can provide some more unusual effective spin-spin interactions. Competition between electronic repulsion and the effective spin-spin interaction with tunable charge density can offer a new platform to show rich unconventional superconductivity phases.

\section{Summary}\label{sec:Summary}

In summary, we have used symmetry arguments and tight-binding models to obtain the coupling between the lattice
dynamics and the electron spin  in adatom decorated graphene. Since the vibration mode transforming according to the $G'$-representation at $K$ and $K'$ and the mode transforming according to $E_1$ at $\Gamma$ are the lowest energy phonons
coupling to the electron-spin, we have focused on them.  The
strength of the phonon to electron-spin coupling constants have been estimated using a tight-binding model. Finally, in order to
study the effect of this coupling on the dynamics of electronic quasi-particle, we have computed the phonon spin-flip rate.
Our results could be relevant for the understanding of the temperature dependence of spin-transport in graphene decorated with heavy adatoms. In particular, the spin-flip rate may be measurable by performing spin-polarize scanning tunneling spectroscopy of the system.

\section{Acknowledgements}

This work was supported by National Center for Theoretical Sciences and Ministry of Science and Technology~(MOST) in Taiwan.  JS You also acknowledges the support from NSC Grant No. 102-2917-I-007-032. MAC also acknowledges support from a start-up fund from National Tsing Hua University.

\appendix
\section{Dynamics matrix of the center force model}\label{a1}
 In this appendix, we provide the details of the derivation of the central-force model introduced in section~\ref{sec:centralforcemodel}
(cf.  Eq.~\eqref{centralforces}). By working
in Fourier space, we obtain
the phonon eigenfrequencies and eigenmodes  of the model. Upon inserting  ${u}_i(\boldsymbol{R})=\frac{1}{\sqrt{N}}\sum_{\boldsymbol{p}} e^{i \boldsymbol{p}\cdot (\boldsymbol{R}+ \boldsymbol{\delta}_{\boldsymbol{r}_j})} u_i(\boldsymbol{p})$  into Eq.~\eqref{centralforces}, the elastic energy
term, $\sum_{ij} \frac{1}{2}k_{ij} (\boldsymbol{\hat{\rho}}_{ij} \cdot(\boldsymbol{u}_i-\boldsymbol{u}_j))^2$, becomes:
\be
\frac{1}{2} \sum_{\boldsymbol{q}} u^{\dagger}(\boldsymbol{q})
\begin{bmatrix} A & 0  & C^*_1 & F^*_1 & C^*_2 & G^*\\
0 & A &F^*_1 & D^*_1 & G^* & D^*_2 \\
C_1 & F_1  & A & 0 & F^*_2 & G\\
F_1 & D_1  &0 & A & G& G^*_2\\
C_2 & G  & F_2 & G^* & B & 0\\
G &  D_2  & G^* & G_2 & 0 & B \end{bmatrix}
         u(\boldsymbol{q}),
 \ee
where
\begin{equation}
u^{T}(\boldsymbol{q}) \equiv\left( x_A(\boldsymbol{q}),  y_A(\boldsymbol{q}),  x_B(\boldsymbol{q}), y_B(\boldsymbol{q}), x_0(\boldsymbol{q}), y_0(\boldsymbol{q}) \right),
\end{equation}
and
\begin{align}
A&=\frac{3}{2}k+\frac{3}{2}k', \\
B&=3k',\\
C_1&=-\frac{3}{4}k[e^{i \boldsymbol{p}\cdot \boldsymbol{b}_{12}}+e^{i \boldsymbol{p}\cdot \boldsymbol{b}_{34}}], \\
C_2 &=-\frac{3}{4}k'[e^{i \boldsymbol{p}\cdot \boldsymbol{\delta}_3}+e^{i \boldsymbol{p}\cdot \boldsymbol{\delta}_5}],\\
D_1&=-\frac{1}{4}k[e^{i \boldsymbol{p}\cdot \boldsymbol{b}_{12}}+e^{i \boldsymbol{p}\cdot \boldsymbol{b}_{34}}+4e^{i \boldsymbol{p}\cdot \boldsymbol{b}_{32}}],\\
D_2&=-\frac{1}{4}k'[e^{i \boldsymbol{p}\cdot \boldsymbol{\delta}_5}+e^{i \boldsymbol{p}\cdot \boldsymbol{\delta}_3}+4e^{i \boldsymbol{p} \cdot \boldsymbol{\delta}_1}],\\
F_1&=-\frac{3}{4}k[e^{i \boldsymbol{p}\cdot \boldsymbol{b}_{12}}-e^{i \boldsymbol{p}\cdot \boldsymbol{b}_{34}}],\\
F_2&=-\frac{3}{4}k'[e^{i \boldsymbol{p}\cdot \boldsymbol{\delta}_2}+e^{i \boldsymbol{p}\cdot \boldsymbol{\delta}_6}]=C_2^*,\\
G&=-\frac{3}{4}k'[e^{i \boldsymbol{p}\cdot \boldsymbol{\delta}_3}-e^{i \boldsymbol{p}\cdot \boldsymbol{\delta}_5}],\\
G_2&=D_2^*.
\end{align}
Here $\boldsymbol{\delta}_{1}=-\boldsymbol{\delta}_{4}\equiv a_0(0,-1),\boldsymbol{\delta}_{3}=-\boldsymbol{\delta}_{6}\equiv a_0(\frac{\sqrt{3}}{2},\frac{1}{2})$ and $\boldsymbol{\delta}_{5}=-\boldsymbol{\delta}_{2}\equiv a_0(-\frac{\sqrt{3}}{2},\frac{1}{2})$ are the vectors joining the adatom to its neighboring carbon atoms;
  $\boldsymbol{b}_{12}=\boldsymbol{b}_{54}\equiv a_0(-\frac{\sqrt{3}}{2},-\frac{1}{2}),$ $\boldsymbol{b}_{32}=\boldsymbol{b}_{56}\equiv a_0(0,1)$ and $\boldsymbol{b}_{34}=\boldsymbol{b}_{16}\equiv a_0(\frac{\sqrt{3}}{2},-\frac{1}{2})$ are the vectors joining a carbon atom on the B sublattice to the neighboring carbons on the A sublattice~(see Fig. \ref{fig_group}).

\section{Perturbation theory}\label{a2}

In this Appendix, we outline the details of the perturbation theory~\cite{Yao,Alicea1} that allows us to integrate out the adatom orbitals and obtain the proximity-induced electron-spin to phonon coupling. The starting electronic Hamiltonian matrix consists of the minimal tight-binding model, Eq.~\eqref{eqHe}, discussed in section~\ref{es} and the electron-phonon coupling, Eq.~\eqref{phononc_c} and Eq.~\eqref{phononc_a}, in section~\ref{sec:md}. By means of a Fourier transformation,
 \be
 d_m(\boldsymbol{R}_I)=\frac{1}{\sqrt{V}}\sum_{\boldsymbol{k}} e^{i \boldsymbol{k}\cdot \boldsymbol{R}_I}d_m(\boldsymbol{k}),\\
   c_{\boldsymbol{r}_j}(\boldsymbol{R}_I)=\frac{1}{\sqrt{V}}\sum_{\boldsymbol{k}} e^{i \boldsymbol{k}\cdot (\boldsymbol{R}_I+\boldsymbol{\delta}_{\boldsymbol{r}_j})}c(\boldsymbol{k}),\\
   u_{\boldsymbol{r}_j}(\boldsymbol{R}_I)=\frac{1}{\sqrt{N}}\sum_{\boldsymbol{k}} e^{i \boldsymbol{k}\cdot (\boldsymbol{R}_I+\boldsymbol{\delta}_{\boldsymbol{r}_j})}u(\boldsymbol{k}),
  \ee
the electronic Hamiltonian matrix in the momentum space can be written as follows:
\be
\hat{H}&=&\sum_{\boldsymbol{k},\boldsymbol{p}}\psi_{\boldsymbol{k}}^{\dagger}H(\boldsymbol{k}, \boldsymbol{k}-\boldsymbol{p})\psi_{\boldsymbol{k}-\boldsymbol{p}},
\ee
where
\be
 H(\boldsymbol{k},\boldsymbol{k}-\boldsymbol{p})&=&\begin{bmatrix}h_g(\boldsymbol{k},\boldsymbol{k}-\boldsymbol{p}) & T(\boldsymbol{k},\boldsymbol{k}-\boldsymbol{p}) \\
 \tilde{T}(\boldsymbol{k},\boldsymbol{k}-\boldsymbol{p}) & h_a(\boldsymbol{k},\boldsymbol{k}-\boldsymbol{p})
 \end{bmatrix},\label{eq:hmatrix}
\ee
and
\begin{equation}
\psi^T_{\boldsymbol{k}}\equiv
\left[c_A(\boldsymbol{k}), c_B(\boldsymbol{k}),d_1(\boldsymbol{k}),d_0(\boldsymbol{k}),d_{-1}(\boldsymbol{k})\right].
\end{equation}
%
Here $h_g(\boldsymbol{k},\boldsymbol{k}-\boldsymbol{p})$ is the Hamiltonian matrix just coming from graphene~(see Eqs.\eqref{tightbinding} and \eqref{phononc_c}), but its explicit form is not important for the calculation sketched below.

$h_a(\boldsymbol{k},\boldsymbol{k}-\boldsymbol{p})$ represents the Hamiltonian of adatoms:
 \begin{small}
 \be
 h_a(\boldsymbol{k},\boldsymbol{k}-\boldsymbol{p})=\begin{bmatrix}
 \epsilon_1+\Lambda_{SO} s^z  & \sqrt{2}\Lambda'_{SO} s^- & 0 \\
 \sqrt{2}\Lambda'_{SO} s^+  & \epsilon_0 & \sqrt{2}\Lambda'_{SO} s^- \\
 0 & \sqrt{2}\Lambda'_{SO} s^+ & \epsilon_1-\Lambda_{SO} s^z
 \end{bmatrix}.\nonumber\\
 \ee
 \end{small}
$T(\boldsymbol{k},\boldsymbol{k}-\boldsymbol{p}) $ and $
\tilde{T}(\boldsymbol{k},\boldsymbol{k}-\boldsymbol{p})$ come from, Eq.~\eqref{c_a} and Eq.~\eqref{phononc_a} respectively, which allow electrons to tunnel between the
adatoms and its neighboring carbons with the influence of adatom's displacement. After some straightforward algebra, we have $T(\boldsymbol{k},\boldsymbol{k}-\boldsymbol{p})\equiv T_0(\boldsymbol{k})\delta_{\boldsymbol{k},\boldsymbol{k}-\boldsymbol{p}}+T_u(\boldsymbol{k},\boldsymbol{k}-\boldsymbol{p}),$ and $\tilde{T}(\boldsymbol{k},\boldsymbol{k}-\boldsymbol{p})\equiv T^{\dagger}_0(\boldsymbol{k})\delta_{\boldsymbol{k},\boldsymbol{k}-\boldsymbol{p}}+\tilde{T}_u(\boldsymbol{k},\boldsymbol{k}-\boldsymbol{p}),$ where $\tilde{T}_u(\boldsymbol{k},\boldsymbol{k}-\boldsymbol{p})$ can be obtained by
  \be
 \tilde{T}_u(\boldsymbol{k},\boldsymbol{k}-\boldsymbol{p})= T^{\dagger}_u(  -\boldsymbol{p}, \boldsymbol{k}-\boldsymbol{p} ).
 \ee
Here we use $ \boldsymbol{p}\rightarrow -\boldsymbol{p}$ first and then $\boldsymbol{k}\rightarrow \boldsymbol{k}-\boldsymbol{p}$ on $T^{\dagger}_u(\boldsymbol{k},\boldsymbol{k}-\boldsymbol{p}).$ $T_0(\boldsymbol{k})$ is defined as:
 \be
 T_0(\boldsymbol{k})&\equiv&\begin{bmatrix} V_1(\boldsymbol{k}) & V_0(\boldsymbol{k}) & V_{-1}(\boldsymbol{k}) \\
 U_1(\boldsymbol{k}) & U_0(\boldsymbol{k}) & U_{-1}(\boldsymbol{k})
 \end{bmatrix},\\
  \ee
  with
  \be
 V_m(\boldsymbol{k})&=&t_{|m|}\sum_{n=1,3,5} e^{i (n-1)\pi/3} e^{-i\boldsymbol{\delta}_n\cdot\boldsymbol{k}},\\
 U_m(\boldsymbol{k})&=&V^*_{-m}(\boldsymbol{k}),
 \ee
 and at the same time
 \begin{small}
  \be
 &&T_u(\boldsymbol{k},\boldsymbol{k}-\boldsymbol{p})\equiv\nonumber\\
 &&\begin{bmatrix}  V_{1A}(\boldsymbol{k},\boldsymbol{k}-\boldsymbol{p}) & V_{0A}(\boldsymbol{k},\boldsymbol{k}-\boldsymbol{p}) & V_{-1A}(\boldsymbol{k},\boldsymbol{k}-\boldsymbol{p}) \\
 V_{1B}(\boldsymbol{k},\boldsymbol{k}-\boldsymbol{p}) & V_{0B}(\boldsymbol{k},\boldsymbol{k}-\boldsymbol{p}) & V_{-1B}(\boldsymbol{k},\boldsymbol{k}-\boldsymbol{p})\\
 \end{bmatrix}\nonumber\\
 \ee
  \end{small}
 where
 \begin{small}
 \be
 &&V_{mA}(\boldsymbol{k}, \boldsymbol{k}-\boldsymbol{p})=\frac{1}{\sqrt{6N}}\frac{\partial t_{|m|}}{\partial l}
\{x_A(\boldsymbol{p})[\frac{\sqrt{3}}{2}e^{i(2\pi/3)m} e^{-i \boldsymbol{\delta}_3\cdot (\boldsymbol{k}-\boldsymbol{p})}\nonumber\\ &-&\frac{\sqrt{3}}{2}e^{i(4\pi/3)m} e^{-i\boldsymbol{\delta}_5\cdot (\boldsymbol{k}-\boldsymbol{p})}]+x_0(\boldsymbol{p})[-\frac{\sqrt{3}}{2}e^{i(2\pi/3)m}e^{-i \boldsymbol{\delta}_3 \cdot \boldsymbol{k}}\nonumber\\
&+&\frac{\sqrt{3}}{2}e^{i(4\pi/3)m} e^{-i\boldsymbol{\delta}_5\cdot \boldsymbol{k}}]+y_A(\boldsymbol{p})[-e^{-i\boldsymbol{\delta}_1\cdot (\boldsymbol{k}-\boldsymbol{p})}\nonumber\\
&+&\frac{1}{2}e^{i(2\pi/3)m}e^{-i\boldsymbol{\delta}_3\cdot (\boldsymbol{k}-\boldsymbol{p})}+\frac{1}{2}e^{i(4\pi/3)m} e^{-i\boldsymbol{\delta}_5\cdot (\boldsymbol{k}-\boldsymbol{p})}]
  \nonumber\\
  &+&y_0(\boldsymbol{p})[+e^{-i\boldsymbol{\delta}_1\cdot \boldsymbol{k}}-\frac{1}{2}e^{i(2\pi/3)m} e^{-i\boldsymbol{\delta}_3\cdot \boldsymbol{k}}-\frac{1}{2}e^{i(4\pi/3)m} e^{-i\boldsymbol{\delta}_5\cdot \boldsymbol{k}}]\}\nonumber\\
 \ee
  \end{small}
  and
   \begin{small}
 \be
  &&V_{mB}(\boldsymbol{k}, \boldsymbol{k}-\boldsymbol{p})=\frac{1}{\sqrt{6N}}\frac{\partial t_{|m|}}{\partial l} \{x_B(\boldsymbol{p})[\frac{\sqrt{3}}{2}e^{i(\pi/3)m} e^{-i\boldsymbol{\delta}_2\cdot (\boldsymbol{k}-\boldsymbol{p})}\nonumber\\
  &-&\frac{\sqrt{3}}{2}e^{i(5\pi/3)m} e^{-i\boldsymbol{\delta}_6\cdot (\boldsymbol{k}-\boldsymbol{p})}]+x_0(\boldsymbol{p})[-\frac{\sqrt{3}}{2}e^{i(\pi/3)m} e^{-i\boldsymbol{\delta}_2 \cdot \boldsymbol{k} }\nonumber\\  &+&\frac{\sqrt{3}}{2}e^{i(5\pi/3)m} e^{-i\boldsymbol{\delta}_6\cdot \boldsymbol{k}} ]+y_B(\boldsymbol{p})[-\frac{1}{2}e^{+i(\pi/3)m} e^{-i\boldsymbol{\delta}_2\cdot (\boldsymbol{k}-\boldsymbol{p})}
  \nonumber\\  &+&e^{i\pi m} e^{-i\boldsymbol{\delta}_4\cdot (\boldsymbol{k}-\boldsymbol{p})}-\frac{1}{2}e^{i(5\pi/3)m} e^{-i\boldsymbol{\delta}_6\cdot (\boldsymbol{k}-\boldsymbol{p})}]\nonumber\\ &+&y_0(\boldsymbol{p})[+\frac{1}{2}e^{i(\pi/3)m} e^{-i\boldsymbol{\delta}_2\cdot \boldsymbol{k}}-e^{i\pi m} e^{-i\boldsymbol{\delta}_4\cdot \boldsymbol{k}}\nonumber\\
  &+&\frac{1}{2}e^{i(5\pi/3)m} e^{-i\boldsymbol{\delta}_6\cdot \boldsymbol{k}}]\}.
 \ee
 \end{small}

To integrate out adatom's degrees of
freedom, we can choose a proper transformation $H\rightarrow \tilde{H}=e^{-S}H e^{S}$ to make $h_g$ and $h_a$ decoupled. When
\be
S=\begin{bmatrix} 0 & M \\
 -M^{\dagger} & 0
 \end{bmatrix},
\ee
 the Hamiltonian matrix is transformed to
\be
\tilde{H}&=&e^{-S}H e^{S}\nonumber\\
 &=&\begin{bmatrix} \tilde{H}_{11} & h_g M+T- M h_a \\
 M^{\dagger} h_g+\tilde{T}-h_a M^{\dagger} & \tilde{H}_{22}
 \end{bmatrix}_{\boldsymbol{k},\boldsymbol{k}-\boldsymbol{p}}.\nonumber\\
\ee
The off-diagonal terms can be eliminated when requiring
\be
\tilde{T}&=&-M^{\dagger} h_g +h_a M^{\dagger},\\
T&=&  M h_a-h_g M,
\ee
or equivalently
\be
M^{\dagger}&=&h_a^{-1} \tilde{T}+ h_a^{-1}M^{\dagger} h_g, \\
M&=& T h_a^{-1}+h_g M h_a^{-1}.
\ee
This transformation yields corrections to $h_g$, such that
\be
\tilde{h}_g&=&h_g-TM^{\dagger}-M\tilde{T}-\frac{1}{2}h_g M M^{\dagger}-\frac{1}{2} M M^{\dagger} h_g\nonumber\\
&+&M h_a M^{\dagger}+O(M^3)\\
&\approx&h_g-Th_a^{-1}\tilde{T},
\ee
where we employ $\frac{1}{2}(-h_g M M^{\dagger}+M h_a M^{\dagger})=\frac{1}{2}T M^{\dagger}$ and $\frac{1}{2}(- M M^{\dagger} h_g+M h_a M^{\dagger})=\frac{1}{2} M \tilde{T}$.
Following the assumption used in Ref.~[\oncite{Alicea1}] that the spin-orbit splitting of adatom orbitals is small compared to the crystal field effects,   we can keep only the leading order terms in $\Lambda'_{SO}$ and $\Lambda_{SO}$ and make $h_a^{-1}\simeq h_0^{-1}-h_0^{-1} h_1 h_0^{-1},$ where $h_a=h_0+h_1$ and $h_0=\textrm{diag}(\epsilon_1,\epsilon_0,\epsilon_1).$ Thus, to the leading order of phonon displacement fields we obtain
\be
 \tilde{h}_g(\boldsymbol{k},\boldsymbol{k}-\boldsymbol{p})_{spin}&\approx& T_0(\boldsymbol{k}) h_0^{-1} h_1  h_0^{-1}\tilde{T}_u(\boldsymbol{k},\boldsymbol{k}-\boldsymbol{p})\nonumber\\&+&T_u (\boldsymbol{k},\boldsymbol{k}-\boldsymbol{p}) h_0^{-1}h_1h_0^{-1}T_0^{\dagger}(\boldsymbol{k}-\boldsymbol{p}).\nonumber \\\label{correction_to_h_g}
 \ee
Here only terms relevant to electron's spin are shown. Eq.~\eqref{correction_to_h_g} can help us to compute strength of the phonon-SO
coupling of Eqs.~(\ref{eq:inter},\ref{eq:intra}).
\section{Diagonalization of the non-interacting electron Hamiltonian}\label{a3}
\wbe
In this appendix, we will show
the electronic Hamiltonian given in Eq.~\eqref{eq:h0},
\be
H_0(\boldsymbol{k})&=&v_F\hbar(\tau_z\sigma_x k_x+\sigma_y k_y)+M \tau_z\sigma_z s_z,
\ee
in the basis,
\be
  \Phi(\boldsymbol{k})= \begin{pmatrix}
  c_{ A,\boldsymbol{K},\uparrow}(\boldsymbol{k}) \\
    c_{ B,\boldsymbol{K},\uparrow}(\boldsymbol{k})\\
   c_{ A,\boldsymbol{K}',\uparrow}(\boldsymbol{k})\\
   c_{ B,\boldsymbol{K}',\uparrow}(\boldsymbol{k})\\
  c_{ A,\boldsymbol{K},\downarrow}(\boldsymbol{k})\\
    c_{ B,\boldsymbol{K},\downarrow}(\boldsymbol{k})\\
   c_{ A,\boldsymbol{K}',\downarrow} (\boldsymbol{k})\\
    c_{ B,\boldsymbol{K}',\downarrow}(\boldsymbol{k})
   \end{pmatrix},
\ee
can be diagonalized by an unitary transformation $\hat{E}(\boldsymbol{k})=U^{-1}H_0(\boldsymbol{k}) U,$ where
the unitary matrix is
\begin{small}
\be
U=\begin{bmatrix}
     0 &  0 & 0  & \frac{M+R}{F_+}& 0 &  0 & 0  & \frac{M-R}{F_-}\\
     0 &   0 & 0 &  \frac{ \hbar vke^{i \theta_{\boldsymbol{k}}}}{F_+} & 0 &   0 & 0 &  \frac{ \hbar vke^{i \theta_{\boldsymbol{k}}}}{F_-}\\
     0 &  0 & \frac{M-R}{F_-} & 0 &0 &  0 & \frac{M+R}{F_+} & 0 \\
     0 &   0 &  \frac{ \hbar v k e^{-i \theta_{\boldsymbol{k}}}}{F_-} & 0 & 0 &   0 &  \frac{ \hbar vke^{-i \theta_{\boldsymbol{k}}}}{F_+} & 0\\
     0 & \frac{-M+R}{F_-}  & 0 & 0 & 0 & \frac{-M-R}{F_+}  & 0 & 0\\
     0 & \frac{ \hbar vke^{i \theta_{\boldsymbol{k}}}}{F_-} & 0 & 0 & 0 & \frac{  \hbar vke^{i \theta_{\boldsymbol{k}}}}{F_+} & 0 & 0\\
    -\frac{M+R}{F_+}       & 0  & 0 & 0 &  -\frac{M-R}{F_-}       & 0  & 0 & 0\\
    \frac{ \hbar vke^{-i \theta_{\boldsymbol{k}}}}{F_+} & 0  & 0 & 0 &   \frac{ \hbar vke^{-i \theta_{\boldsymbol{k}}}}{F_-} & 0  & 0 & 0
   \end{bmatrix},
\ee
\end{small}
\wee
with $\tan\theta_{\boldsymbol{k}}=k_y/k_x,$
 $
R(\boldsymbol{k})\equiv\sqrt{M^2+\hbar^2 v^2 ( k_x^2+ k_y^2)},$ $F_+(\boldsymbol{k})\equiv\sqrt{(M+R(\boldsymbol{k}))^2+\hbar^2v^2 ( k_x^2+ k_y^2)} $ and
  $
 F_-(\boldsymbol{k})\equiv\sqrt{(M-R(\boldsymbol{k}))^2+\hbar^2v^2 ( k_x^2+ k_y^2)}.
 $
  After this unitary transformation, the electronic Hamiltonian in the grand canonical ensemble becomes
a diagonal energy eigenvalue matrix,
\begin{small}
\be
\hat{E}(\boldsymbol{k}) -\mu&=&\begin{bmatrix}  (R(\boldsymbol{k})-\mu) I_{4\times 4} &  0 \\
 0 &  (-R(\boldsymbol{k})-\mu) I_{4\times 4} \end{bmatrix}\\&=&\begin{bmatrix}  \varepsilon_{\boldsymbol{k},+} I_{4\times 4} &  0 \\
 0 &  \varepsilon_{\boldsymbol{k},-} I_{4\times 4} \end{bmatrix},
\ee
\end{small}
where $\varepsilon_{\boldsymbol{k},\pm}=\pm\sqrt{M^2+\hbar^2 v^2 \boldsymbol{k}^2}-\mu$. Here $\mu$ is the chemical potential and $I_{4\times 4}$ is the $4\times 4$ unit matrix.
\section{The spin-flip rate}\label{a4}
To compute the spin-flip rate, we can straightforwardly select contribution from the following four terms:
 $(U^{\dagger}{L}^{\alpha}U)_{(a,\uparrow),(b,\downarrow)}$ from Eq. \eqref{eq:FermiGorden}, where $a,b=(\pm,K/ K')$. Therefore the spin-flip rate of a quasiparticle with $\eta=(\pm,K (K'), \uparrow(\downarrow))$ quantum numbers can be written as
 \wbe
 \begin{small}
\be
\Gamma^{\mathrm{flip}}_{\boldsymbol{k},\pm}
&=&\frac{2 \pi}{\hbar}\sum_{\boldsymbol{p}}\frac{2 A_{K}^2}{ 2\omega_{K}}\left\{g_{+}(\omega_{K},\boldsymbol{k},\boldsymbol{p})+g_{-}(\omega_{K},\boldsymbol{k},\boldsymbol{p})\pm \frac{-g_{+}(\omega_{K},\boldsymbol{k},\boldsymbol{p})+g_{-}(\omega_{K},\boldsymbol{k},\boldsymbol{p})}{R(\boldsymbol{k}) R(\boldsymbol{k}-\boldsymbol{p})}\left[  M^2-\frac{1}{2} k |\boldsymbol{k}-\boldsymbol{p}|v^2 \cos(\theta_{\boldsymbol{k}}-\theta_{\boldsymbol{k}-\boldsymbol{p}}) \right]\right\}\notag\nonumber\\
&+&\frac{2 \pi}{\hbar}\sum_{\boldsymbol{p}}\frac{ C_{\Gamma}^2}{ 2\omega_{\Gamma}}\left\{ g_{+}(\omega_{\Gamma},\boldsymbol{k},\boldsymbol{p})+g_{-}(\omega_{\Gamma},\boldsymbol{k},\boldsymbol{p})\pm \frac{-g_{+}(\omega_{\boldsymbol{\Gamma}},\boldsymbol{k},\boldsymbol{p})+g_{-}(\omega_{\Gamma},\boldsymbol{k},\boldsymbol{p})}{R(\boldsymbol{k})R(\boldsymbol{k}-\boldsymbol{p})}
\left[ M^2+k |\boldsymbol{k}-\boldsymbol{p}|v^2 \cos(\theta_{\boldsymbol{k}}-\theta_{\boldsymbol{k}-\boldsymbol{p}}) \right] \right\},\label{eq:d1}
\ee
 \end{small}
where
 \begin{small}
\be
g_{\pm}(\omega_{\alpha},\boldsymbol{k},\boldsymbol{p})&=&\frac{1}{N}\left[1+n(\omega_{\alpha})-f(\varepsilon_{\boldsymbol{k}-\boldsymbol{p},\pm})\right]\delta(\omega-(\varepsilon_{\boldsymbol{k}-\boldsymbol{p},\pm}+\omega_{\alpha}))
+\frac{1}{N}\left[n(\omega_{\alpha})+f(\varepsilon_{\boldsymbol{k}-\boldsymbol{p},\pm})\right]\delta(\omega-(\varepsilon_{\boldsymbol{k}-\boldsymbol{p},\pm}-\omega_{\alpha})),\label{gpm}
\ee
 \end{small}
 \wee
and $R(\boldsymbol{k})\equiv\sqrt{M^2+\hbar^2 v^2 ( k_x^2+ k_y^2)}.$
Because the band structures are
symmetric between unlike
spins and unlike valleys at low energy, the spin-flip rate from the quasiparticle of channel $\eta$ is dependent on the energy of quasiparticles, $\omega=\varepsilon_{\boldsymbol{k},\pm},$ but independent on the spin and valley labels.
In the following calculation we would like to deal with the summation on the internal momentum $\boldsymbol{p}.$ By rewriting the summation as integral over the internal momentum $\boldsymbol{p},$ Eq. \eqref{eq:d1} becomes
\wbe
 \begin{small}
\be
\Gamma^{\mathrm{flip}}_{\boldsymbol{k},\pm}
&=&\frac{2 \pi}{\hbar}\sum_{\lambda=\pm}\frac{\Omega}{ (2\pi)^2}\int^{\infty}_0 dp  \int^{2 \pi}_0 d\theta  \frac{ 2 A_{K}^2}{2 \omega_{K}}p\left\{g_{\lambda}(\omega_{K},\boldsymbol{k},\boldsymbol{p})\mp \frac{{\lambda }g_{\lambda}(\omega_{K},\boldsymbol{k},\boldsymbol{p})}{R(\boldsymbol{k}) R(\boldsymbol{k}-\boldsymbol{p})}\left[ M^2-\frac{1}{2}k v^2 (k-p \cos \theta) \right]\right\}\notag\nonumber\\
&+&\frac{2 \pi}{\hbar}\sum_{\lambda=\pm}\frac{\Omega}{ (2\pi)^2}\int^{\infty}_0 dp  \int^{2 \pi}_0 d\theta \frac{ C_{\Gamma}^2}{ 2\omega_{\Gamma}}p\left\{ g_{\lambda}(\omega_{\Gamma},\boldsymbol{k},\boldsymbol{p})\mp \frac{{\lambda}g_{\lambda}(\omega_{\boldsymbol{\Gamma}},\boldsymbol{k},\boldsymbol{p})}{R(\boldsymbol{k})R(\boldsymbol{k}-\boldsymbol{p})}
\left[  M^2+kv^2 (k-p \cos \theta)  \right]\right\}\\
&=&\frac{2 \pi}{\hbar}  2 A_{K}^2\sum_{\lambda=\pm}\frac{2 \Omega}{ (2\pi)^2}\int^{\infty}_0 dp  \int^{-1}_1 d y  \frac{ 1}{ 2\omega_{K}}\frac{-p}{\sqrt{1-y^2}}\left\{g_{\lambda}(\omega_{K},\boldsymbol{k},\boldsymbol{p})\mp \frac{{\lambda }g_{\lambda}(\omega_{K},\boldsymbol{k},\boldsymbol{p})}{R(\boldsymbol{k}) R(\boldsymbol{k}-\boldsymbol{p})}\left[ M^2-\frac{1}{2} k v^2 (k-p y)\right]\right\}\notag\nonumber\\
&+&\frac{2 \pi}{\hbar}C_{\Gamma}^2\sum_{\lambda=\pm}\frac{2 \Omega}{ (2\pi)^2}\int^{\infty}_0 dp  \int^{-1}_1 d y \frac{ 1}{ 2\omega_{\Gamma}}\frac{-p}{\sqrt{1-y^2}}\left\{ g_{\lambda}(\omega_{\Gamma},\boldsymbol{k},\boldsymbol{p})\mp \frac{{\lambda}g_{\lambda}(\omega_{\boldsymbol{\Gamma}},\boldsymbol{k},\boldsymbol{p})}{R(\boldsymbol{k})R(\boldsymbol{k}-\boldsymbol{p})}\left[ M^2+kv^2 (k-p y)  \right]\right\},\label{eq:d4}
\ee
 \end{small}
\wee
where $\cos(\theta_{\boldsymbol{k}}-\theta_{\boldsymbol{k}-\boldsymbol{p}})=(k-p \cos \theta)/|\boldsymbol{k}-\boldsymbol{p}|,$ $y=\cos\theta$ and the azimuthal angle part is replaced as $ d\theta=\frac{-1}{\sqrt{1-y^2}}dy$. Let us  introduce the function:
\be
e_{\boldsymbol{k}-\boldsymbol{p},\lambda}(y)&=&\lambda R(\boldsymbol{k}-\boldsymbol{p})\nonumber\\ &\equiv& \lambda\sqrt{M^2+\hbar^2v^2(k^2+p^2-2kp y )},
\ee with $\lambda=\pm$. Thus, the delta function of Eq.~\eqref{gpm} becomes
\begin{small}
\be
\delta(\omega'-(\frac{e_{\boldsymbol{k}-\boldsymbol{p},\lambda}(y)}{\hbar}\pm \omega_{\alpha}))= \frac{ |e_{\boldsymbol{k}-\boldsymbol{p},\lambda}(y_{\mp,\alpha})| }{\hbar v^2 kp}\delta(y-y_{\mp,\alpha})\nonumber\\
\times\Theta(\lambda\hbar(\omega'\mp \omega_{\alpha})-M)
\ee
\end{small}
where $\omega'\equiv\omega+\mu$ and
\begin{small}
\be
y_{\mp,\alpha}=\frac{-\hbar^2(\omega'\mp\omega_{\alpha})^2+M^2+\hbar^2v^2(k^2+p^2)}{2\hbar^2v^2kp}.
\ee
\end{small}
Furthermore the interval of the integral of $p$ is limited by  $-1\leq y_{\mp,\alpha} \leq 1$~(remember $y=\cos\theta$). From the inequality,
$y_{\mp,\alpha}\leq 1,$ we obtain
\be
 p_{1,\mp,\alpha}\leq p \leq p_{2,\mp,\alpha}
\ee
where
\be
p_{1,\mp,\alpha}=k-\sqrt{\frac{\hbar^2(\omega'\mp\omega_{\alpha})^2-M^2}{v^2 \hbar^2}}
 \ee
 and
 \be
 p_{2,\mp,\alpha}=k+\sqrt{\frac{\hbar^2(\omega'\mp\omega_{\alpha})^2-M^2}{v^2 \hbar^2}}\\\nonumber
 \ee
with $\sqrt{\hbar^2(\omega'\mp\omega_{\alpha})^2-M^2}$ being real.
From the other inequality, $-1\leq y_{\mp,\alpha},$ we obtain
\be
p \leq p^*_{2,\mp,\alpha},  p^*_{1,\mp,\alpha}\leq p
\ee
where
\be
p^*_{1,\mp,\alpha}=-k+\sqrt{\frac{\hbar^2(\omega'\mp\omega_{\alpha})^2-M^2}{v^2 \hbar^2}}
\ee
and
\be
p^*_{2,\mp,\alpha}=-k-\sqrt{\frac{\hbar^2(\omega'\mp\omega_{\alpha})^2-M^2}{v^2 \hbar^2}}.
\ee
Therefore these inequalities limit the range of $p$ within $\textrm{Max}\{p_{1,\mp,\alpha},p^*_{1,\mp,\alpha}\}\leq p \leq p_{2,\mp,\alpha}.$
\wbe
Bringing results of above discussions into Eq. \eqref{eq:d4}, we have~(skip coefficients $\frac{2 \pi}{\hbar}  2 A_{K}^2$ and $\frac{2 \pi}{\hbar}  2 C_{\Gamma}^2$ here)
\begin{small}
\be
&&
\sum_{\lambda=\pm}\frac{2\Omega}{ (2\pi)^2}\int^{\infty}_0 dp  \int^{-1}_1 dy\frac{-p}{\sqrt{1-y^2}}\frac{1}{ 2 \omega_{\alpha}}g_{\lambda}(\omega_{\alpha},\boldsymbol{k},\boldsymbol{p})\\
&=&\frac{2\Omega}{N (2\pi)^2}\sum_{\lambda=\pm}\int^{\infty}_0 dp  \int^{-1}_1 dy\frac{-p}{\sqrt{1-y^2}}\frac{1}{ 2 \hbar v^2kp\omega_{\alpha}}\Bigg[\Big|e_{\boldsymbol{k}-\boldsymbol{p},\lambda}(y_{-,\alpha})\Big|\big(1+n(\omega_{\alpha})-f(e_{\boldsymbol{k}-\boldsymbol{p},\lambda}(y)-\mu)\big)\delta(y-y_{-,\alpha})\Theta\big(\lambda\hbar(\omega'- \omega_{\alpha})-M\big)\nonumber\\
&&+\Big|e_{\boldsymbol{k}-\boldsymbol{p},\lambda}(y_{+,\alpha})\Big|\big(n(\omega_{\alpha})+f(e_{\boldsymbol{k}-\boldsymbol{p},\lambda}(y)-\mu)\big)\delta(y-y_{+,\alpha})\Theta\big(\lambda\hbar(\omega'+ \omega_{\alpha})-M\big)\Bigg]\\
&=&\frac{2 \Omega}{ N v^2 k (2 \pi)^2}\sum_{\lambda=\pm}\int^{\infty}_0 dp \Bigg\{ \Big[1+n(\omega_{\alpha})-f(\lambda\hbar|\omega'-\omega_{\alpha}|-\mu)\Big]\frac{|\omega'-\omega_{\alpha}| \Theta(1-|y_{-,\alpha}|) }{2 \omega_{\alpha}\sqrt{1-(y_{-,\alpha})^2}} \Theta\big(\lambda\hbar(\omega'- \omega_{\alpha})-M\big)\nonumber\\&&
 + \Big[n(\hbar\omega_{\alpha})+f(\lambda\hbar|\omega'+\omega_{\alpha}|-\mu
)\Big]\frac{|\omega'+\omega_{\alpha}| \Theta(1-|y_{+,\alpha}|) }{2 \omega_{\alpha} \sqrt{1-(y_{+,\alpha})^2}} \Theta\big(\lambda\hbar(\omega'+ \omega_{\alpha})-M\big)\Bigg\},
\ee
\end{small}
and
\begin{small}
\be
&&
\sum_{\lambda=\pm}\frac{2\Omega}{ (2\pi)^2}\int^{\infty}_0 dp  \int^{-1}_1 dy\frac{-p}{\sqrt{1-y^2}}\frac{1}{ 2 \omega_{{K}}}\frac{{\lambda }g_{\lambda}(\omega_{{K}},\boldsymbol{k},\boldsymbol{p})}{R(\boldsymbol{k}) R(\boldsymbol{k}-\boldsymbol{p})}\left[ M^2-\frac{1}{2}k v^2 (k-p y) \right]\\
&=&\frac{2\Omega}{N (2\pi)^2}\sum_{\lambda=\pm}\int^{\infty}_0 dp  \int^{-1}_1 dy\frac{-p}{\sqrt{1-y^2}}\frac{{\lambda }}{ 2 \hbar v^2k p R(\boldsymbol{k}) \omega_{{K}}}\Bigg\{\Big[1+n(\omega_{{K}})-f(e_{\boldsymbol{k}-\boldsymbol{p},\lambda}(y)-\mu)\Big]\delta(y-y_{-,\alpha})\Theta\big(\lambda\hbar(\omega'- \omega_{{K}})-M\big)\nonumber\\
&&+\Big[n(\omega_{{K}})+f(e_{\boldsymbol{k}-\boldsymbol{p},\lambda}(y)-\mu)\Big]\delta(y-y_{+,\alpha})\Theta\big(\lambda\hbar(\omega'+ \omega_{{K}})-M\big)\Bigg\}\left[ M^2-\frac{1}{2} k v^2 (k-p y) \right]\\
&=&\frac{2 \Omega}{ N \hbar v^2 k R(\boldsymbol{k})  (2 \pi)^2}\sum_{\lambda=\pm}{\lambda }\int^{\infty}_0 dp \Bigg\{ \Big[1+n(\omega_{{K}})-f(\lambda\hbar|\omega'-\omega_{{K}}|-\mu)\Big]\frac{ \Theta(1-|y_{-,\alpha}|)(2 M^2-k^2 v^2 +pk v^2 y_{-,\alpha} ) }{4 \omega_{{K}}\sqrt{1-(y_{-,\alpha})^2}}\nonumber\\&&\times \Theta\big(\lambda\hbar(\omega'- \omega_{{K}})-M\big)
+ \Big[n(\hbar\omega_{{K}})+f(\lambda\hbar|\omega'+\omega_{{K}}|-\mu
)\Big]\frac{ \Theta(1-|y_{+,\alpha}|)(2 M^2-k^2 v^2 +pk v^2 y_{+,\alpha} ) }{4 \omega_{{K}} \sqrt{1-(y_{+,\alpha})^2}} \Theta\big(\lambda\hbar(\omega'+ \omega_{{K}})-M\big)\Bigg\},\nonumber\\
\ee
\end{small}
\wee
where $|e_{\boldsymbol{k}-\boldsymbol{p},\lambda}(y_{\pm,\alpha})|=|\omega'\pm\omega_{\alpha}|. $
In our case the optical phonon mode $\omega_{\alpha}$ does not have momentum dependence~(to the leading order),
so we just need to compute the integral:
\begin{small}
\be
&&\int^{\infty}_0 dp \frac{ \Theta(1-|y_{\pm,\alpha}|) }{\sqrt{1-(y_{\pm,\alpha})^2}}\\
&=&\int^{p_{2,\pm,\alpha}}_{\textrm{Max}\{p_{1,\pm,\alpha},p^*_{1,\pm,\alpha}\}} dp \frac{ 1 }{\sqrt{1-(y_{\pm,\alpha})^2}}\\
&=&\int^{p_{2,\pm,\alpha}}_{\textrm{Max}\{p_{1,\pm,\alpha},p^*_{1,\pm,\alpha}\}} dp \frac{ p }{\sqrt{p^2-(\rho_{\pm,\alpha}+\beta p^2)^2}}\\
&=&\int^{p_{2,\pm,\alpha}^2}_{p_{1,\pm,\alpha}^2} dx \frac{1 }{2\sqrt{x-(\rho_{\pm,\alpha}+\beta x)^2}}= 2 \pi k,\label{integral1}
\ee
\end{small}
and
\be
&&\int^{\infty}_0 dp \frac{p y_{\pm,\alpha} \Theta(1-|y_{\pm,\alpha}|) }{\sqrt{1-(y_{\pm,\alpha})^2}}\\
&=&\int^{p_{2,\pm,\alpha}^2}_{p_{1,\pm,\alpha}^2} dx \frac{\rho_{\pm,\alpha}+\beta x }{2\sqrt{x-(\rho_{\pm,\alpha}+\beta x)^2}}= 2 \pi k^2, \label{integral2}\\\nonumber
\ee
where $y_{\pm,\alpha}\equiv\frac{\rho_{\pm,\alpha}+\beta p^2}{p},$ $\rho_{\pm,\alpha}\equiv\frac{-\hbar^2(\omega'\pm\omega_{\alpha})^2+M^2+\hbar^2v^2k^2}{2\hbar^2v^2k},$ $\beta\equiv\frac{1}{2k},$ and $x\equiv p^2.$ Because of Eq. \eqref{integral1} and \eqref{integral2}, $\cos(\theta_{\boldsymbol{k}}-\theta_{\boldsymbol{k}-\boldsymbol{p}})$ in Eq. \eqref{eq:d1} gives no contribution to the spin flip rate.
Therefore terms associated with $\omega_{K}$ in $\Gamma^{\mathrm{flip}}_{\boldsymbol{k},\pm}$ can be simplified as
\wbe
\begin{small}
\begin{align}
&\sum_{\lambda=\pm}\frac{2\Omega}{ (2\pi)^2}\int^{\infty}_0 dp  \int^{-1}_1 dy\frac{-p}{\sqrt{1-y^2}}\frac{1}{ 2\omega_{K}}\Bigg\{g_{\lambda}(\omega_{K},\boldsymbol{k},\boldsymbol{p})\mp\frac{{\lambda }g_{\lambda}(\omega_{K},\boldsymbol{k},\boldsymbol{p})}{R(\boldsymbol{k}) R(\boldsymbol{k}-\boldsymbol{p})}\left[ M^2-\frac{1}{2}k v^2 (k-p y) \right]\Bigg\}\\
&=\frac{\Omega}{2 N  \hbar \omega_{K} \pi v^2 }
\Bigg\{
\Big[ 1+n(\omega_{K})-f(\omega-\omega_{K})\Big]\Big[ \hbar \big|\omega'-\omega_{K}\big|\mp{\textrm{sgn}(\omega'-\omega_{K}) } \frac{ M^2}{  \hbar R(\boldsymbol{k}) }\Big] \Theta\big(\hbar^2(\omega'-\omega_{K})^2-M^2\big)\nonumber\\
&\quad  + \Big[ n(\hbar\omega_{K})+f(\omega+\omega_{K})\Big]\Big[ \hbar\big|\omega'+\omega_{K}\big|\mp{\textrm{sgn}(\omega'+\omega_{K})}\frac{ M^2}{  \hbar R(\boldsymbol{k}) }\Big]\Theta\big(\hbar^2(\omega'+\omega_{K})^2-M^2\big)\Bigg\}\\
&\equiv L_{\pm}(\omega,\omega_{K})\label{eq:flip_of_omegaK};
\end{align}
Terms associated with $\omega_{\Gamma}$ can be simplified as
\begin{align}
&\sum_{\lambda=\pm}\frac{2\Omega}{ (2\pi)^2}\int^{\infty}_0 dp  \int^{-1}_1 dy\frac{-p}{\sqrt{1-y^2}}\frac{1}{ 2 \omega_{\Gamma}}\Bigg\{g_{\lambda}(\omega_{\Gamma},\boldsymbol{k},\boldsymbol{p})\mp\frac{{\lambda }g_{\lambda}(\omega_{\Gamma},\boldsymbol{k},\boldsymbol{p})}{R(\boldsymbol{k}) R(\boldsymbol{k}-\boldsymbol{p})}\Big[ M^2+k v^2 (k-p y) \Big]\Bigg\}= L_{\pm}(\omega,\omega_{\Gamma}).\label{eq:flip_of_omegaGamma}
\end{align}
\end{small}.
\wee
 Since the energy of incoming quasi-particle is $\omega,$ we can set $R(\boldsymbol{k})=|\omega'|=|\omega+\mu|$  in Eq.~\eqref{eq:flip_of_omegaK} and \eqref{eq:flip_of_omegaGamma} and drop the $\boldsymbol{k}$ dependence of $\Gamma^{\mathrm{flip}}_{\boldsymbol{k},\pm}(\omega).$
To ensure incoming quasi-particles within bands, $\Theta(\pm\omega'-M)$ is introduced with $L_{\pm}(\omega,\omega_{\alpha})$ and $\textrm{sgn}(\omega'\pm\omega_{K,\Gamma})$ can be simplified as $\textrm{sgn}(\omega')$~($M>\hbar\omega_{K,\Gamma}$ in our case). Thus
 the spin-flip rate of a spin polarized quasi-particle at a valley in a band denoted by $\lambda=\pm$ is obtained to be
 \wbe
\be
\Gamma^{\mathrm{flip}}_{\lambda}(\omega)&=&\Gamma^{\mathrm{flip}}_{\mathrm{inter},\lambda}(\omega)+\Gamma^{\mathrm{flip}}_{\mathrm{intra},\lambda}(\omega)
=\frac{2 \pi}{\hbar}\big[2{ A_K^2}L_{\lambda}(\omega,\omega_K)+{ C_{\Gamma}^2}L_{\lambda}(\omega,\omega_{\Gamma})\big],
\ee
where
\be
L_{\lambda}(\omega,\omega_{\alpha})&=&\Theta(\lambda\hbar\omega'-M)\frac{\Omega}{2  N  \hbar \omega_{\alpha} \pi v^2 }  \Bigg\{
\Big[1+n(\omega_{\alpha})-f(\omega-\omega_{\alpha})\Big]
 \Big[ \hbar \big|\omega'-\omega_{\alpha}\big|-\frac{ M^2}{  \hbar |\omega' |}\Big]  \Theta (\hbar^2(\omega'-\omega_{\alpha})^2-M^2)  \notag\\
&&+  \Big[n(\omega_{\alpha})+f(\omega+\omega_{\alpha})\Big] \Big[ \hbar \big|\omega'+\omega_{\alpha}\big|-\frac{ M^2}{  \hbar| \omega'| }\Big] \Theta(\hbar^2(\omega'+\omega_{\alpha})^2-M^2)\Bigg\}.
\ee
\wee
In the expression of $\Gamma^{\mathrm{flip}}_{\lambda}(\omega),$ the scattering event associated with $\omega_K$ phonons of $G'$ modes is identified as inter-valley spin-flip  $\Gamma^{\mathrm{flip}}_{\mathrm{inter},\lambda}(\omega),$ whereas the event associated with $\omega_{\Gamma}$ phonons of $E_1$ modes is identified as intravalley spin-flip $\Gamma^{\mathrm{flip}}_{\mathrm{intra},\lambda}(\omega).$


\begin{thebibliography}{99}
%
\bibitem{Kane}
C. L. Kane and E. J. Mele, Phys. Rev. Lett. {\bf 95}, 226801~(2005).
%
\bibitem{Hasan}
M. Z. Hasan and C. L. Kane, Rev. Mod. Phys. {\bf 82}, 3045~(2010).
%
\bibitem{Qi}
X.-L. Qi and S.-C. Zhang, Rev. Mod. Phys. {\bf 83}, 1057~(2011).
%
\bibitem{mercury_tell}
M. K\"onig, S. Wiedmann, C. Br\"une, A. Roth, H. Buhmann, L.~W.
Molenkamp, X.-L. Qi, S.C. Zhang, Science {\bf 308}, 766~(2007).
%
\bibitem{other}
I. Knez,  R.-R. Du, G. Sullivan, Phys. Rev. Lett. {\bf 107}, 136603~(2011).
%
\bibitem{novoselov}
K. S. Novoselov, A. K. Geim, S. V. Morozov, D. Jiang, Y. Zhang, S. V. Dubonos, I. V. Grigorieva, A. A. Firsov, Science {\bf 306}, 666~(2004).
%
\bibitem{neto}
A. H. Castro Neto, F. Guinea, N. M. R. Peres, K. S. Novoselov, and A. K. Geim, Rev. Mod. Phys. {\bf 81}, 109~(2009).
%
\bibitem{Hernando}
D. Huertas-Hernando, F. Guinea, and A. Brataas, Phys.
Rev. B {\bf 74}, 155426~(2006).
%
\bibitem{Yao}
Y. Yao, F. Ye, X.-L. Qi, S.-C. Zhang, and Z. Fang, Phys. Rev. B {\bf 75}, 041401~(2007).
%
\bibitem{neto2}
 A. H. Castro Neto, and  F. Guinea, Phys. Rev. Lett. {\bf 103}, 026804~(2009).
%
\bibitem{natphys1}
J. Balakrishnan, G.-K. Wai Koon, M. Jaiswal,
A. H. Castro Neto, B. \"Ozyilmaz, Nat. Phys. {\bf 9},
284~(2013).
%
\bibitem{Min}
H. Min, J. E. Hill, N. A. Sinitsyn, B. R. Sahu, L. Kleinman,
and A. H. MacDonald, Phys. Rev. B {\bf 74}, 165310~(2006).
%
\bibitem{Gmitra}
M. Gmitra, S. Konschuh, C. Ertler, C. Ambrosch-Draxl,
and J. Fabian, Phys. Rev. B {\bf 80}, 235431~(2009).

%
\bibitem{Samir}
 Samir Abdelouahed, A. Ernst, J. Henk, I. V. Maznichenko,
and I. Mertig, Phys. Rev. B {\bf 82}, 125424~(2010).
%
\bibitem{Qiao}
Z. Qiao, S.~A. Yang, W. Feng,
W.-K. Tse, J. Ding, Y. Yao, Jian Wang, and Qian Niu, Phys. Rev. B {\bf 82}, 161414~(2010).
%
\bibitem{Alicea1}
Conan Weeks, Jun Hu, Jason Alicea, Marcel Franz, and R. Wu, Phys. Rev. X {\bf 1}, 021001~(2011).
%
\bibitem{Jun}
J. Hu, J. Alicea, and R. Wu, M. Franz, Phys. Rev. Lett. {\bf 109}, 266801~(2012).
%
\bibitem{Katsnelson}
M. I. Katsnelson, \emph{
Graphene: Carbon in Two Dimensions}, Cambridge University Press~(Cambridge, UK, 2012).
%
\bibitem{jaja2}
J. Balakrishnan, G.~K. Wai Koon1, Ahmet Avsar, Y. Ho, J.~H. Lee, M. Jaiswal, S.-J. Baeck, J. H. Ahn, A. Ferreira, M.~A. Cazalilla, A. H. Castro Neto, and Barbaros \"Ozyilmaz, Nat. Comm. {\bf 5}, 4748~(2014).

%
\bibitem{prl2014}
A. Ferreira, T. Rappoport, and M. A. Cazalilla, A. H. Castro Neto,
Phys.~Rev.~Lett.~{\bf 102} 066601~(2014).
%
\bibitem{Yu2014}
Z. Jia~\emph{et al}, report arXiv:1409.8090 (2014).
%
\bibitem{Rashba}
D. Marchenko, \emph{et al.}, Nat. Commun. {\bf 3}, 1232 (2012).
%
\bibitem{Hector}
F, Calleja \emph{et al.},  Nat. Phys. {\bf 11}, 43 (2014).
%
\bibitem{Basko}
D.~M. Basko, Phys. Rev. B {\bf 78}, 125418~(2008).
%
\bibitem{Guinea0}
F. Guinea, J. Phys. C: Condens. Matt. {\bf 14}, 3345~(1981).
%
\bibitem{Castro}
A. H. Castro Neto and F. Guinea, Phys. Rev. B {\bf 75}, 045404~(2007).
%
\bibitem{Edward}
E. McCann and V. I. Fal’ko, Phys. Rev. Lett. {\bf 108}, 166606~(2012).

\bibitem{ochoa}
H. Ochoa, A. H. Castro Neto, V. I. Fal'ko, and F. Guinea, Phys. Rev. B {\bf 86}, 245411~(2012).
%
\bibitem{private}
R. Ribeiro, \emph{private communication} (2013).
%

\bibitem{coleman}
P. Coleman, Lecture notes  \emph{Introduction to Many-body Physics}.
\bibitem{kawakami}
 K. Pi, W. Han, K. M. McCreary, A. G. Swartz, Y. Li, and R. K. Kawakami, Phys. Rev. Lett. {\bf 104}, 187201 (2010).
\bibitem{Niu}
H. Jiang, Z. Qiao, H. Liu, J. Shi, and Q. Niu, Phys. Rev.
Lett. {\bf 109}, 116803 (2012).
\bibitem{Paul}
P. Soul\'{e} and M. Franz, report arXiv:1402.6638 (2014).
\bibitem{Roche}
A. Cresti \emph{et al}.,  Physical Review Letters {\bf 113}, 246603 (2014).
%
\bibitem{unpub}
J.-S. You, D.-W. Wang, and M. A. Cazalilla, \emph{to be published}.
%
\bibitem{Guinea}
F. Guinea and B. Uchoa, Phys. Rev. B {\bf 86}, 134521~(2012).
%
\bibitem{Roldan}
R. Roldan, E. Cappelluti and F. Guinea, Physical Review B {\bf 88}, 054515~(2013).
%
\end{thebibliography}
\end{document}